\documentclass[12pt]{iopart}

\usepackage{iopams}  
\usepackage{hyperref}
\usepackage{graphicx}
\usepackage{subcaption}
\usepackage{iopams}
\usepackage{dcolumn}
\usepackage{bm}
\expandafter\let\csname equation*\endcsname\relax
\expandafter\let\csname endequation*\endcsname\relax
\usepackage{amsmath}
\usepackage[left=1in, right=1in, top=1in, bottom=1in]{geometry}
\usepackage{verbatim}
\usepackage{titling}
\usepackage{parskip}
\usepackage{array}
\usepackage{color}
\usepackage{latexsym}
\usepackage{amsthm}
\usepackage{amssymb}
\DeclareGraphicsExtensions{.jpg,.pdf, .mps, .png, .eps, .ps, .EPS,.gif}

\newcommand{\od}[2]{\frac{d #1}{d #2}}

\def\bc{\begin{center}}
\def\ec{\end{center}}
\def\bea{\begin{eqnarray}}
\def\eea{\end{eqnarray}}
\newcommand{\avg}[1]{\langle{#1}\rangle}

\begin{document}

\title{Beyond COVID-19: Network science and  sustainable exit strategies}

\author{J. Bell$^{1}$,G. Bianconi$^{1,2}$,D. Butler $^{1}$,J. Crowcroft $^{3}$,P. C. W. Davies$^4$, C. Hicks$^{1}$, H. Kim$^{4,5,6}$,I.Z. Kiss$^7$, F. Di Lauro$^7$,C. Maple$^{1,8}$,A. Paul$^{9,10}$,M. Prokopenko$^{11,12}$, P. Tee$^4$ and S. Walker$^4$ }

\address{$^1$ The Alan Turing Institute, 96 Euston Rd, London NW1 2DB, United Kingdom. \{jbell,chicks,dbutler\}@turing.ac.uk}\vspace{-0.3cm}
\address{$^2$ School of Mathematical Sciences, Queen Mary University of London, London, E1 4NS, United Kingdom. g.bianconi@qmul.ac.uk}\vspace{-0.3cm}
\address{$^3$ University of Cambridge, Cambridge, CB2 1TN, United Kingdom. jon.crowcroft@cl.cam.ac.uk}\vspace{-0.3cm}
\address{$^4$ Beyond Center for Fundamental Concepts in Science, Arizona State University, Tempe, Arizona 85287 USA. \{paul.davies, sara.i.walker, hkim78, ptee2\}@asu.edu}\vspace{-0.3cm}
\address{$^5$ ASU-SFI Center for Biosocial Complex Systems, Arizona State University and Santa Fe Institute, USA}\vspace{-0.3cm}
\address{$^6$ School of Earth and Space Exploration, Arizona State University, Tempe, AZ, USA}\vspace{-0.3cm}
\address{$^7$ Department of Mathematics, University of Sussex, Falmer, Brighton BN1 9QH, United Kingdom. \{I.Z.Kiss, F.Di-Lauro\}@sussex.ac.uk }\vspace{-0.3cm}
\address{$^8$ University of Warwick, Coventry CV4 7AL, United Kingdom.  CM@warwick.ac.uk}\vspace{-0.3cm}
\address{$^{9}$ DESY, Notkestra{\ss}e 85, D-22607 Hamburg, Germany. apaul2@alumni.nd.edu}\vspace{-0.3cm}
\address{$^{10}$ Institut f\"ur Physik, Humboldt-Universit\"at zu Berlin, D-12489 Berlin, Germany}\vspace{-0.3cm}
\address{$^{11}$ Centre for Complex Systems, The University of Sydney, Australia. mikhail.prokopenko@sydney.edu.au}\vspace{-0.3cm}
\address{$^{12}$ The Marie Bashir Institute for Infectious Diseases and Biosecurity, The University of Sydney, Australia.}\vspace{-0.3cm}

\ead{ptee2@asu.edu (Correspondence address)}

\vspace{10pt}
\begin{indented}
\item[] \today
\end{indented}

\begin{abstract}
On May $28^{th}$ and $29^{th}$, a two day workshop was held virtually, facilitated by the Beyond Center at ASU and Moogsoft Inc.
The aim was to bring together leading scientists with an interest in Network Science and Epidemiology to attempt to inform public policy in response to the COVID-19 pandemic.
Epidemics are at their core a process that progresses dynamically upon a network, and are a key area of study in Network Science.
In the course of the workshop a wide survey of the state of the subject was conducted.
We summarize in this paper a series of perspectives of the subject, and where the authors believe fruitful areas for future research are to be found.
\end{abstract}

%
%
\section{Introduction}
\subsection{Background and Aims of the Workshop}

Reports of a new viral infection with lethal and pandemic potential emerged in the Wuhan province of China in December of 2019 \cite{Wu2020}.
It was clear from early reports that this new virus had severe respiratory complications and could have alarming fatality rates.
The virus, officially designated SARS-CoV-2 by the International Committee on Taxonomy of Viruses, proceeded to create a localized epidemic in Wuhan, resulting in a severe lock down in an attempt to control the outbreak.
This has subsequently proven to have not controlled the epidemic of the virus, now  commonly known as COVID-19 or Corona Virus, and the world is faced with the first widespread pandemic since the `Spanish Flu' outbreak of 1918.

\begin{figure}[t]
	\centering
     	\includegraphics[scale=0.4]{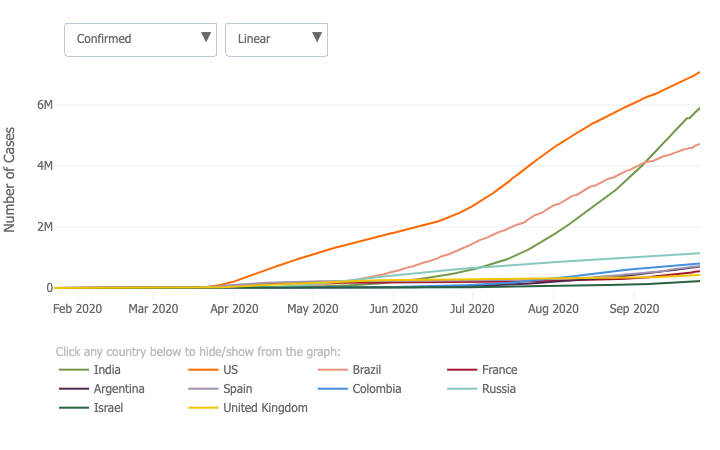}
     	\caption{Cases of COVID-19 for the countries with the top 10 fatality rates. \cite{JHU2020}}
      	\label{fig:OutbreakData}
\end{figure}

The pathology of the disease is significantly higher than seasonal flu \cite{Berlin2020}, and severe symptoms can terminate in a fatal cytokine release resulting in severe inflammatory response, high fever, hypoxia and eventually death.
Initial indications from studies of the outbreak in Wuhan indicate that this fatality rate is not evenly spread demographically, with case fatality rates amongst individuals 80 years and over being estimated between $10\%$ to $18\%$, whereas rates at half that age are barely $1\%$ \cite{Verity2020}.
Regardless, it is clear that COVID-19 is a deadly disease, and as of July $7^{th}$, it has claimed $551,686$ victims, and the initial demographics of morbidity are however likely to change as the pandemic progresses.
Indeed there is strong evidence from the 1918 Spanish Flu that although the first wave of the infection had enhanced mortality in elderly people, the second and subsequent waves killed more indiscriminately \cite{Chowell2010} indicating that it is possible that the future progress of COVID-19, should there be a second wave, could have greater impact on the young than hitherto.
In Fig. \ref{fig:OutbreakData} we reproduce statistics from the Johns Hopkins Coronavirus Resource Center \cite{JHU2020} for the ten countries with the largest current outbreaks.
We note that the data on a linear scale underlines that the spread of the disease is accelerating, and we may be at the beginning rather than substantially into the pandemic.

Public policy response to the pandemic has been a dramatic economic and societal lockdown, essentially plunging the affected nations into a kind of deep freeze.
As evidence emerged that transmission of the virus was mediated by respiratory exhalation of `droplets' or, via contact, social distancing was imposed forcing millions to isolate or shelter in place.
The economic impact of this policy has been dramatic.
US government data on the effect on the Chinese economy \cite{Malden2020} provides an illustration of what even the short lived lock down of one region produced, including a $21.2\%$ drop in retail sales for the {\sl whole} of China, and, a drop in the use of public transport from $70$-$80$ million trips per day to around $10$ million, as the lockdown was imposed.
Around the western world the impact has begun to be felt with for example, the United Kingdom seeing a decline in GDP of $20.4\%$ in the month of April alone \cite{Scruton2020}, and some commentators pointing to an economic recession that will be the most severe in recorded history.

On the face of it, those in charge of public policy are presented with a choice between public health and public wealth, although in reality the two objectives are likely interlinked in complex and unknown ways.
There is, however, a common and perhaps overlooked link between the progression of the disease and the operation of society and the economy.

Underlying the transmission of a disease requiring contact, or at least proximity, is a network of contacts.
Similarly, societal activity requires the mobility of citizens, whether it is to places of work, education, leisure or the international travel necessary to facilitate the business of business.
It is possible, perhaps, to regard the choice between economic activity and restricting the progression of the disease as an optimization problem.
The dynamics of epidemics on networks has been well studied \cite{newman2002spread,Kiss2017}, and transport networks, for example, have been well studied using techniques of network science \cite{Albert2002}.
Given that the transmission networks for the disease may well be the same as the network over which activity is conducted, a simple question naturally arises.
Is it possible to leverage the well understood dynamic properties of networks and epidemics thereupon to propose schemes for the partial lifting of lockdown restrictions that would maximally enable the normal prosecution of an individuals daily life whilst restricting the spread of the pandemic?
As a motivational example of where this is routinely done, one of the authors has successfully applied techniques from the field of network science to maximize the operational activity of an impaired communications network \cite{Tee2017}.

It was for this reason that on May $28^{th}$ and $29^{th}$, a two day virtual workshop was convened by the Beyond Center at Arizona State University.
The workshop united experts from across a range of disciplines, with network science being a common thread.
In the paper we seek to summarize the perspectives that were shared at this workshop, and point to where the participants felt the most promising areas of study currently lie.
As such, this paper hopes to provide a perspective upon the state of the understanding of this extremely important problem and motivate further research upon the issues outlined within.

\subsection{Outline of this Paper}

This paper is a curation of summary reports authored by the speakers at the workshop.
In order to provide a coherent overview of the discussion we have organized the paper into three major parts.
We begin with an overview of the theoretical network models in Sections \ref{sec:tee} and  \ref{sec:kiss}.
Here we cover the latest techniques that are being used to analyze how the standard epidemic models of such as SIR/SIS are enhanced by the considerations of the structure of contact networks.
It can be shown that this has a dramatic effect on the disease induced herd immunity (DIHI) threshold, and has important consequences for the management of the pandemic. 
Specifically the structural nature of the contact network and the details of any actions such as lockdowns impacts significantly the procession of the disease.

In Section \ref{sec:bianconi}  we discusses novel theoretical questions raised by the modeling of COVID-19. We will reveal the role of stochasticity and criticality  in plateauing time series, which provides an important additional source of uncertainties too often neglected in modeling frameworks.
Moreover we indicate that network science is the most suitable tool to theoretically investigate the efficiency of the track and  tracing apps at the system level. In fact the network science approach  allows to capture non-linear effects of the spreading dynamics.

In Section \ref{sec:prokopenko} we turn our attention to another important modeling technique, Agent Based Modeling (ABM), which has been successfully applied to the initial progress of the epidemic in Australia.
This approach permits the detailed use of scenario planning to model, for example, the effects of social distancing measures. 
By essentially creating a digital `twin' of the epidemic this technique accurately predicted many of the features of the early progress of the disease, incorporating many complex features of the demographics and geography of the country.

Key to the management of the epidemic has been the use of so-called `track and trace', whereby the contacts of infected individuals are identified and isolated to prevent further infection.
In Section \ref{sec:crowcroft} we give an overview of the technology of contact tracing applications. 
In this section, we discuss the requirements for identity systems to support these apps, and the way that there are common design patterns for how the apps and the identity systems themselves can be implemented.
This has important consequences for various properties of the systems in terms of protection of privacy with regards to identity, and attributes (``is infected'') associated with that identity.

In Section \ref{sec:kimpaul} a theoretical analysis of the effectives of contact tracing approaches to epidemic control is explored.
Building upon a probabilistic model  the conclusion is that participation in such a control approach needs to be widespread, although the precise degree of participation is still not a settled question.
What is clear, and discussed in this section, is that socio-economic factors play an important role in the progress of the disease.
Key to understanding this is the development of advanced algorithms to understand more deeply the distribution of immune individuals (those that have recovered from the disease), and we assess in this section Machine Learning approaches to this, and also how ABMs can be used to understand lockdown reversal strategies.

We end with concluding remarks in Section \ref{sec:conclusion}.

\section{Network Science and Epidemiology}
\label{sec:tee}

\subsection{Non-Network Compartmental Epidemic Models}
The quantitative analysis of epidemiology dates back to the work of Kermack and McKendrick \cite{kermack1927contribution,kermack1932contributions,kermack1933contributions}, almost a century ago.
Their work evolved into the familiar ``compartmental" models, which explore the dynamics of a population whose individuals move between the ``compartments" or categories of infections.
There are a number of  variants of  such models that use different compartments, and different schema for the transition of an individual from between them.
As an illustration we will briefly describe one of the better known variants, the SIR model.
Here the compartments are characterized by the time-series of the numbers of individuals in the population that are susceptible to infection (S), are infected (I) and  have recovered  from the disease or  have died (R).
The time evolution of the population in each of the compartments is therefore time-dependent, and encodes the dynamics of the epidemic. The evolution of the epidemics  may be  described  by a system of Ordinary Differential Equations (ODEs) that can be solved to determine the progress of the disease.

A central, and very important, assumption of the SIR model is that the population mixes homogeneously.
This means that the probability that a randomly chosen member of the population is infected (moves from $S \rightarrow R$) is uniform, which greatly simplifies the analysis.
Furthermore, an individual in the susceptible population can encounter and be infected by any member of the infected population,
and the only factors that impact the dynamics are the relative sizes of these populations.
This explicitly excludes finer details such as location factors, where in a particular area susceptible individuals may  be present at lower density compared to the average. 

For a population of total size $N$,  the system may be analyzed with  three coupled ODEs \cite{barabasi2016network,Kiss2017} 

\begin{align}\label{eqn:SIR}
    \od{S(t)}{t} &= -\beta I(t) \frac{S(t)}{N} \text{,} \\
    \od{I(t)}{t} &= \beta I(t)\frac{S(t)}{N} -\gamma I(t) \text{,} \\
    \od{R(t)}{t} &= \gamma I(t) \text{.}
\end{align}
Here $S(t)$,$I(t)$,and $R(t)$ represent the time varying numbers of the population in each compartment, $\beta$  represents the rate a susceptible person is infected  and $\gamma$ the rate at which an infected individual either recovers or dies. 
There will be $\beta I(t)$  infecting encounters in unit time with the $S(t)/N$ portion of the population available to be infected, and a number  $\gamma I(t)$  of infected individuals either recover and become immune to infection or die.
In Section \ref{sec:kiss} these equations are treated in more depth, and the reader should note that there the symbol $\tau$ is used in place of $\beta$.
The ratio of the two parameters $R_0=\frac{\beta}{\gamma}$, is known as the basic reproductive ratio, and it is widely known by the general public  as the key measure of the severity of the spreading of the disease.
A simple interpretation of $R_0$ is the average number of susceptible persons that are infected by an infected individual. 
Therefore  large values of $R_0$ are associated with public policy responses such as lockdowns, and the ubiquitous references to exponential growth and doubling times.
Indeed by inspection of Eqs. \eqref{eqn:SIR} we can see that for $R_0 > 1$, that $I(t)$ increases with time, and conversely for $R_0 < 1$ it decreases. 
For $R_0>1$ the epidemics is in the {\em supercritical regime} and the epidemics outbreak will affect a finite fraction of the population. For $R_0<1$ the epidemics is in the {\em subcritical regime} and the epidemics outbreak dies out affecting only an infinitesimal fraction of the population. For $R_0=1$ the epidemics is in the {\em critical regime} and is the epidemics starts with a single infected individual it will infect an infinitesimal fraction of the population however the epidemics will be affected strongly by stochastic effect and the size of the outbreak is difficult to predict and can have very wide fluctuations.

The simplistic SIR model is problematic for many reasons, not least of which is that it is not possible in general to assess the value of $R_0$ at a given point in time; it is generally inferred retrospectively using historical data. 
Indeed the model assumes that $R_0$ as defined as the ratio of $\beta$ to $\gamma$ is fixed in time, and of course as measures are taken to combat an epidemic this is not generally true.
Instead we can refer to $R$ as a point in time reproduction rate, which is possible to determine in many ways using historical data.
One such method is to use macroscopic parameters such as the doubling time and the average duration $t_c$ between an individual becoming infected and going on to infect another person. 
This requires certain assumptions regarding the distribution of macroscopic parameters, and to an extent the underlying epidemic model, but it allows for an empirical estimation of the value of $R$ \cite{wallinga2007generation}.
For example, if we define $r$ as the growth rate of an infection in a unit time (number of new cases per day for example), in the most naive linear model $R=1+rt_c$, which relies upon the assumption that $t_c$ is constant.
Alternatively is we apply a simple multiplicative growth model, it is possible to show that

\begin{equation}
    R=\exp{ (r t_c) } \text{.}
\end{equation}
The values of $R$ obtained using these two different models can be very different.
This hopefully serves to illustrate that the actual value of $R$ is very much dependent upon assumptions implicit in the underlying model.
In short its use is problematic in isolation as a guide to the severity and progress of an epidemic.

\subsection{Introducing Networks into Epidemic Models}

Infections require a physical mechanism of transmission between infected and susceptible individuals. This does not have to be physical contact (as in the case of sexually transmitted diseases), but can arise when individuals are within close enough proximity for sufficient time that an airborne pathogen can pass from one individual to another. The precise biological mechanism for the transmission of the disease is assumed to be encoded in the parameters of model operating on these networks.

One is naturally led to the concept of contact networks, where one represents individuals (or places) as nodes, and the links represent individuals coming into contact. Starting with a number of infected nodes the epidemic then proceeds by transmission via the links according to a certain transmission probability. Clearly the structure of this network can have a profound effect upon the progress of the disease, as the degree of a node (number of links) represents the number of people available to be infected by that individual. When the phrase “breaking the chain of infection” is used, this translates into isolating infected nodes by severing these links, being the basis of quarantine measures.

Models of real world networks are extremely well studied, and it has been known for some time that many of them possess the ``small world'' property \cite{watts1998collective}, whereby non-locality in the connectivity of nodes in the network creates ‘shortcuts’ in the network, which has an important effect on the propagation of an epidemic. 

An important property of real world networks, such as contact graphs, is their degree distribution $P(k)$ of nodes, which quantifies the probability of a randomly selected node having $k$ links. Real world networks are often scale-free, i.e. they have a power-law degree distribution of the 
form $P(k)\propto k^{-\alpha}$,  with values of $\alpha$ 
typically in the range $2.0<\alpha<3.0$ \cite{barabasi2016network}.
Examples range from citation graphs to transport networks, however other social networks such as the mobile-phone contact indicating close friends might display much larger power-law exponent $\alpha\simeq 8$ \cite{onnela2007structure} while contacts in schools, hospital or workplaces might have a more homogeneous degree distribution \cite{isella2011s,stehle2011high,vanhems2013estimating}.

In a much cited result, the power law distribution was shown to be a natural outcome of the preferential attachment model \cite{Albert2002}, where networks evolve by new nodes linking preferentially to ‘popular’ or high degree nodes. There are of course many different randomly-generated networks, but the success of the preferential attachment hypothesis was the clear connection it provided between a physical model of network growth and the measured properties of real networks.

In spite of the overall success of network models applied to the real world, the adaptation of such models to epidemics is in fact non-trivial, and can lead to some surprising results \cite{Satorras_epidemic_revmod_2015,barrat2008dynamical}. For example, consider the  application of scale-free networks to epidemiology, which began with the so-called {\em degree class} approximation of Pastor-Satorras and Vespignani  \cite{pastor2001epidemic}.

This approach assumes that all nodes of a given degree are statistically equivalent, and applies a variant of the SI model (nodes are designated to be either susceptible or infected) to each cohort of nodes sharing the same degree. It is possible to solve these equations using an underlying assumption of the degree distribution, and obtain an expression for the characteristic time for the spread of the infection $\tau_{SI}$ in terms of the moments of $k$:

\begin{equation}
    \tau_{SI} = \frac{\langle{k}\rangle}{ \beta(\langle k^2 \rangle - \langle k \rangle) } \text{.}
\end{equation}
The value $\tau_{SI}$ enters into the dynamics as a time-scale factor.
If $i_k$ is the fraction of nodes of degree $k$ that are infected, it can be shown that  ${di_k}/{dt} \propto k e^{t/\tau_{SI}}$.
As $\langle k^2 \rangle \rightarrow \infty$ for scale free graphs this has the very surprising consequence that the characteristic spreading time for the disease is zero. In essence, the disease propagates extremely fast affecting rapidly a large fraction of the network, because of the presence of ‘hubs’ in the network, which allow an infected individual to rapidly infect a large fraction of  network of contacts.

The SIR model can be mapped to  link percolation which can then be used to predict the expected size of an outbreak as a function of the {\em trasmissibility} of the disease $T$\cite{moore2000epidemics,newman2002spread}. 
In this approach the   nodes of the network represent single individuals and the links  represent the social contacts of an individuals providing the possible routes  for the transmission of the disease. The transmission of the diseases from an infected to a neighbour  susceptible individual occurs with a probability $T$ called 'transmissibility', which is dependant upon the length of time an individual is infectious $\tau$, and the rate $\beta$ at which an individual infects one  of its  contact.
Assuming that the rate $\beta$ is independent on time,  the transmissibility can be expressed as   
\begin{equation}
T=1-e^{- \beta \tau}.
\end{equation}
Therefore the SIR outbreak can be represented as a network in which the route of transmission is indicated by  the  ``occupied' links',  where each link is occupied with probability $T$.
Therefore the   sub-graph connected only the occupied links, indicates the infected cluster of individuals. As a function of $T$ this cluster can be very small, affecting an infinitesimal fraction of all the node of the networks or very large, or {\em giant} if it includes a finite fraction of all the nodes. In this latter case the infected cluster corresponds to the giant component (GC) of the network \cite{dorogovtsev2008critical}.
The GC is the  connected sub-graph of a network which includes a finite fraction of all the nodes. The probability of a random node being a member of the GC does not evolve smoothly, but instead exhibits critical behavior describing a  phase transition as a function of the control parameter $T$.  For $T$ smaller than the critical value $T_c$  the largest connected cluster involves only an infinitesimal fraction of the nodes of the network, implying that for these value of the transmissibility the epidemic dies out before becoming widely spread in the population. However for $T$ larger than the critical value $T_c$ a GC emerges indicating that the outbreak  affects a finite fraction of the population (here modelled by its corresponding contact network).

Very interestingly there is an important effect of the degree distribution on the critical properties of the SIR model.
The critical value $T_c$ of the transmissibility , called the {\em epidemic threshold}, on a random network with given degree distribution is given by \cite{Satorras_epidemic_revmod_2015,barrat2008dynamical}
\begin{equation}
T_c=\frac{\avg{k}}{\avg{k^2}-\avg{k}}.
\end{equation}
This result implies that  if the network has a homogeneous degree distribution, with a finite second moment of the degree distribution (i.e. with  $\avg{k^2}<\infty$) the { epidemic threshold} $T_c$ is finite.
Therefore if  the transmissibility $T$ is smaller that the epidemic threshold $T_c$ the outbreaks does not affects significantly the population.
However for scale-free networks having diverging second moment of the degree distribution (i.e. with  $\avg{k^2}\to \infty$) the epidemic threshold vanishes
$T_c\to 0$
as the number $N$ of the nodes of the network diverge, i.e. $N\to\infty$. This means that also epidemics with very small transmissibility can become pandemics.
This result implies  that  globalized societies are very prone to pandemics as the air-travel connections are well known to be described by scale-free networks. 

An important network property of disease spreading is the fact that the nodes of the network are not equally likely to get the infection. In particular nodes of high degree are more likely to get the infection of nodes with lower degree. Therefore it comes as no surprise that people with many contacts including politicians and bus drivers have been more likely to be infected in the first wave of COVID-19.

This result is rather intuitive as if we assume that each connection of an individual is equally likely to be route for transmission of the disease, an individual with more connections will be more likely to get the disease than an individual with less connections.

\subsubsection{Herd Immunity and the Friendship Paradox}
The Herd Immunity (HI) threshold is the percentage of the population which, if immune, prevents the number of infected individuals  from growing  due to the scarcity of susceptible individuals.
A classic result from non-network models \cite{smith1970prospects} states HI in terms of $R_0$ as 

\begin{equation}
    h=1-\frac{1}{R_0} \text{.}
\end{equation}
This is straightforward to derive by noting that for compartmental models a non-growing epidemic requires that $R_0(1-h)=1$. 
With an estimate of $R_0$ for COVID-19 of between $2.0 - 3.0$, we arrive at a value of $h$ between $50\%$ and $66\%$ needed for the epidemic to peak in absence of containment measures.
This is a very important value as the lower it is, the more likely that the epidemic will naturally abate, but for reference at the time of writing only $2\%$ of US population is currently infected.

In addition to HI, there is a related concept Disease Induced Herd Immunity (DIHI) that can occur as the natural result of the first wave of infection, and is not dependent upon intervention measures.
In essence this relies upon the disease spreading quickly through the more highly connected nodes early in the epidemic, such that if any intervention such as lockdown is eased the network of contacts is sufficiently disrupted to prevent further spread.

Using the mapping of the SIR model to percolation, it is rather intuitive that the removal of links or nodes from the contact graph can  reduce the GC through which the epidemic preferentially progresses, and would at a certain point prevent the growth of the epidemic. This removal of links or nodes is represented in the public health domain by immunization or social distancing measures. In fact immunized individuals or individuals in quarantine are nodes of the network that cannot spread the epidemics any more. Critically, for scale-free networks, the behavior of the size of the GC as nodes are removed is  different from other random networks  \cite{albert2000error,dorogovtsev2008critical}. Targeted measures that remove nodes of high degree would collapse the GC more quickly than would random removal (by immunization or isolation).

The foregoing important insight is readily understandable in terms of the so-called friendship paradox, first introduced by Scott Feld  \cite{feld1991your}. It follows from the simple observation that on average your friends have more friends that you do. Because your ‘friendly friends’ have more contacts they are more likely to be immunized by targeted immunization, leading to an improved efficiency of targeted immunization than of random immunization in scale-free networks. 

Note however that the efficiency of this targeted immunization on a scale-free network  strongly depends on two parameters, the power-law exponent $\alpha$ of the power-law degree distribution and the minimum degree of the nodes. 
Let us  assume $T=1$ let us consider  a scale-free network with degree distribution $P(k)=Ck^{-\alpha}$ with $m\leq k\leq 1000$. For   $\alpha=2.5$ and minimum degree $m=2$ in order to suppress the epidemics it  is necessary to perform a targeted immunization over a fraction of a population given by $f\simeq 20\%$ but this fraction increases to $f\simeq 45\%$ for $m=4$ and $f\simeq 59\%$ for $m=6$. Therefore the efficiency of the targeted immunization strategy is very sensitive to the minimum degree in the network (corresponding to the minimum number of connections of the nodes in the population).

Assuming that the epidemics gives a sufficiently long immunity, different waves of COVID-19 might act as effective targeted immunization of the population. Indeed since the first wave will affect more likely the hubs of the network, the first wave might act as targeted  immunization \cite{newman2005threshold}. However the shielding effect of this self-immunization phenomenon depends strongly on the network topology and in particular on the minimum degree of the nodes.
Moreover this self-immunization phenomenon and the implied threshold for herd immunity depends sensitively on the presence of heterogeneity, a topic that we take up in Section \ref{sec:kiss}.

Using these models, estimates of DIHI range from 20\% to 60\%, with the lower range having profound implications for public policy if it were to be the applicable to COVID-19. It should however be stressed that these approaches involve toy models, and rely upon the transmission network being scale free, which is a questionable assumption. Consider, for example, the effect of public transport. The daily commute brings together disjoint sub-populations in conditions ideal for disease transmission. This has the effect of homogenizing the contact graphs of the commuters and altering any prediction of DIHI based upon scale freedom.

In conclusion it is evident that the fine details of the contact graphs have an important role in the dynamics of epidemics. However, small changes in the structure of these graphs can have a dramatic effect on the predictions for DIHI with obvious and important consequences for public policy. As such their use in isolation is probably not advisable, but they are nevertheless an important tool in the management of epidemics. In the next two sections we will consider further important consequences of contact networks and the models that are built upon them.

\section{Overview of Network Based Epidemic Models}
\label{sec:kiss}

The transmission of a disease depends not only on the intrinsic characteristics of the pathogen that causes it, but, equally importantly, on the network of disease-transmitting contacts within the population. If this contact structure is ignored and homogeneous random mixing is assumed then it is well known that if a fraction $1-1/\mathcal{R}_0$ of the population cannot be infected (e.g. vaccinated preventively or already infected) then the residual susceptible population can no longer sustain an epidemic. Instead, a recent observation~\cite{Britton2020} is that, by taking into account heterogeneities in the population (to be understood in the broadest sense but here captured as heterogeneities in contacts), this threshold can be crossed when far fewer individuals have been infected. This is because the disease acts like a targeted vaccine, preferentially `immunizing' higher-risk individuals who play a greater role in transmission.  Therefore, a controlled `first wave' may leave behind a residual population that can no longer sustain a `second wave' once interventions are lifted. This concept of exploiting heterogeneities is often overlooked but it has a profound impact on the outcome of an epidemic and should be considered for policy making decisions. In light of this, we systematically analyze a number of well-known mean-field models for networks and we consider the question of herd-immunity induced by a disease spreading on networks with different characteristics.

Mean-Field models play a major role in the formulation of epidemic models on Networks, with many results following  from the mathematical analysis of these~\cite{Porter2016,pastor2015epidemic,Chen2014,Holme2017}.
Such models might not inform policy making directly, but they remain useful to gain insights on the epidemic process itself,  while remaining mathematically tractable. 
Before investigating the impact of contact heterogeneities on herd immunity levels, we briefly describe the mean-field models that we use. We order models by their relative complexity, corresponding to gradually incorporating more features of the underlying network. 

\subsection{Exact equations}
The exact ODE system for the SIR model describes the evolution for the expected number of nodes in given statuses. A formal derivation from the exact system is given in~\cite{Kiss2017}. The system of equation is:
\begin{eqnarray}
 \dot{[S]} &=& -\tau [SI], \nonumber \\
 \dot{[I]} &=& \tau [SI]  - \gamma[I],\nonumber \\
 \dot{[R]} &=& \gamma[I],\nonumber \\ \label{mean-field}
\end{eqnarray}
where $[SI]$ is the expected number of links between susceptible and infected nodes. This system is not closed: to solve it we need an expression for the evolution of the expected number of $S-I$ links (expected number of pairs), which in turn requires us to describe the system at the level of triples, and so on. Mean-field models curtail this expansion at some level, by expressing higher order quantities in terms of lower order ones. These methods are also known as  closures since they lead to a self-consistent system of differential equations. Such closure are often found by taking into account the network structure up to a certain level (e.g. mean-degree, degree distribution and clustering for example).
\subsection{Homogeneous Mean-field model}
The simplest mean-field model incorporates solely the average degree $\langle k \rangle$ of the network, and it can be described by the following equations
\begin{eqnarray}
 \dot{[S]} &=& -\tau \frac{\langle k \rangle}{N} [S][I], \nonumber \\
 \dot{[I]} &=& \tau \frac{\langle k \rangle}{N} [S][I]  - \gamma[I],\nonumber \\
 \dot{[R]} &=& \gamma[I].\nonumber \\
\label{hom-mean-field}
\end{eqnarray}
The intuition behind this closure is to consider all the nodes as having the same degree $\langle k \rangle$. There are $[S]$ susceptible nodes with $\langle k \rangle [S]$ stubs connecting them to their neighbors. We assume that infected nodes are distributed randomly on the network, so that the probability that a neighbor is infected is $\frac{[I]}{N}$. The expression for $[SI]$ is then given by
\[
[SI] \sim \langle k \rangle [S] \frac{[I]}{N}.
\]

\subsection{Degree-based mean-field model}
In the homogeneous mean-field model ~\eqref{hom-mean-field}, the system is closed by two approximations: each node has the same degree and infected nodes are uniformly distributed on the network. The degree-based mean-field model (also called Heterogeneous Mean Field~\cite{pastor2015epidemic}) improves the closure by removing the first approximation, i.e. by incorporating the degree distribution in the system. We denote with $[S]_k(t)$ the expected number of susceptible nodes with degree $k$ at time $t$, similarly for $[I]_k$ and $[R]_k$. We define $[S] = \sum_{k=1}^\infty [S]_k$, similarly $[I]$ and $[R]$. As in the homogeneous mean-field model, we average the infection pressure across all the infected nodes. The resulting ODEs are
\begin{eqnarray}
 \dot{[S_k]} &=& -\tau k [S_k] \pi_I, \nonumber \\
 \dot{[I_k]} &=& \tau k [S_k] \pi_I - \gamma[I_k],\nonumber \\
 \dot{[R_k]} &=& \gamma[I_k],\nonumber \\
\pi_I &=& \frac{\sum_{\ell=1}^M \ell [I_\ell]}{\sum_{\ell=1}^M \ell N_\ell}, \label{eq:degree-based_mean-field}
\end{eqnarray}
where $N_\ell=P_{n,p}(\ell)N$ is the number of nodes with degree $\ell$, and $P_{n,p}(l)$ is the negative binomial or versions of it used later to simulate degree distributions with different heterogeneities. This system keeps track of the degree distribution and hence the heterogeneity in it, but mixing between nodes of different degrees happens at random but proportionally to degree~\cite{pastor2015epidemic,Kiss2017}. In  reality, correlations build up:  if a node has a neighbor infected, it is more likely to be infected itself than if it were  randomly  selected  among the susceptible nodes. Hence, better models are required, which results in closures at the level of triples.

\subsection{Heterogeneous pairwise model}

In the heterogeneous pairwise model, we also consider the expected number of links connecting a node of degree $k$ in state $A$ to a node of degree $\ell$ in state $B$~\cite{house2011insights,Kiss2017}, that is $[A_kB_\ell]$. To include these quantities into anode system, we need to write an expression for triples of the form $[A_kB_\ell C_m]$. By doing so, we are effectively including the correlations between nodes in the same state. The closure is done at the level of triples (i.e. triples are approximated by singles and pairs), and hence an approximation for the triples are needed. In this framework we also consider clustering (i.e. the propensity with which two neighbors of a node are themselves connected). The closure in this case is:
\begin{equation}
[A_kB_\ell C_m] = \frac{\ell-1}{\ell}\left((1-\varphi)\frac{[A_kB_\ell][B_\ell C_m]}{[B_j]} + \varphi\frac{[A_kB_\ell][B_\ell C_m][C_mA_k]}{[A_k][B_\ell][C_m]}\right),    
\label{eq:triple-closure}
\end{equation}
where $\varphi$ is the global clustering coefficient in the network. For the un-clustered case we simply set $\varphi=0$. The derivation of this can be found in~\cite{house2011insights}, for example.
The resulting ODEs are,
\begin{eqnarray}
 \dot{[S_k]} &=& -\tau \sum_{\ell}[S_kI_{\ell}], \nonumber \\
 \dot{[I_k]} &=& \tau \sum_{\ell}[S_kI_{\ell}] - \gamma[I_k],\nonumber \\
 \dot{[R_k]} &=& \gamma[I_k],\nonumber \\
    \dot{[S_kI_\ell]} &=& - \gamma[S_kI_\ell] + \tau \left(\sum_{\alpha}[S_kS_\ell I_{\alpha}] -\sum_{\alpha}[I_{\alpha}S_kI_\ell] -[S_kI_\ell] \right), \nonumber \\
 \dot{[S_kS_\ell]} &=& - \tau\left( [S_kS_\ell I] + [I S_k S_\ell] \right),  \nonumber\\
  \dot{[I_kI_\ell]} &=& -2\gamma[I_kI_\ell]+ \tau\left( \sum_{\alpha}[I_{\alpha}S_kI_\ell] + \sum_{\alpha}[I_kS_\ell I_{\alpha}]+[S_kI_\ell]+[I_kS_\ell] \right), 
 \label{eq:pairwise}
\end{eqnarray}
where triples are closed using equation~\eqref{eq:triple-closure}. This system includes t both the  full degree distribution of the network and the evolution of the $[SI]$ pairs. The number of equations in the heterogeneous pairwise model grows very large if the network has degrees of many different types (since there is an equation for $\dot{[S_kI_\ell]}$ for every $k$, $\ell$ pair). 

\begin{figure}
\centering
\includegraphics[scale=0.95]{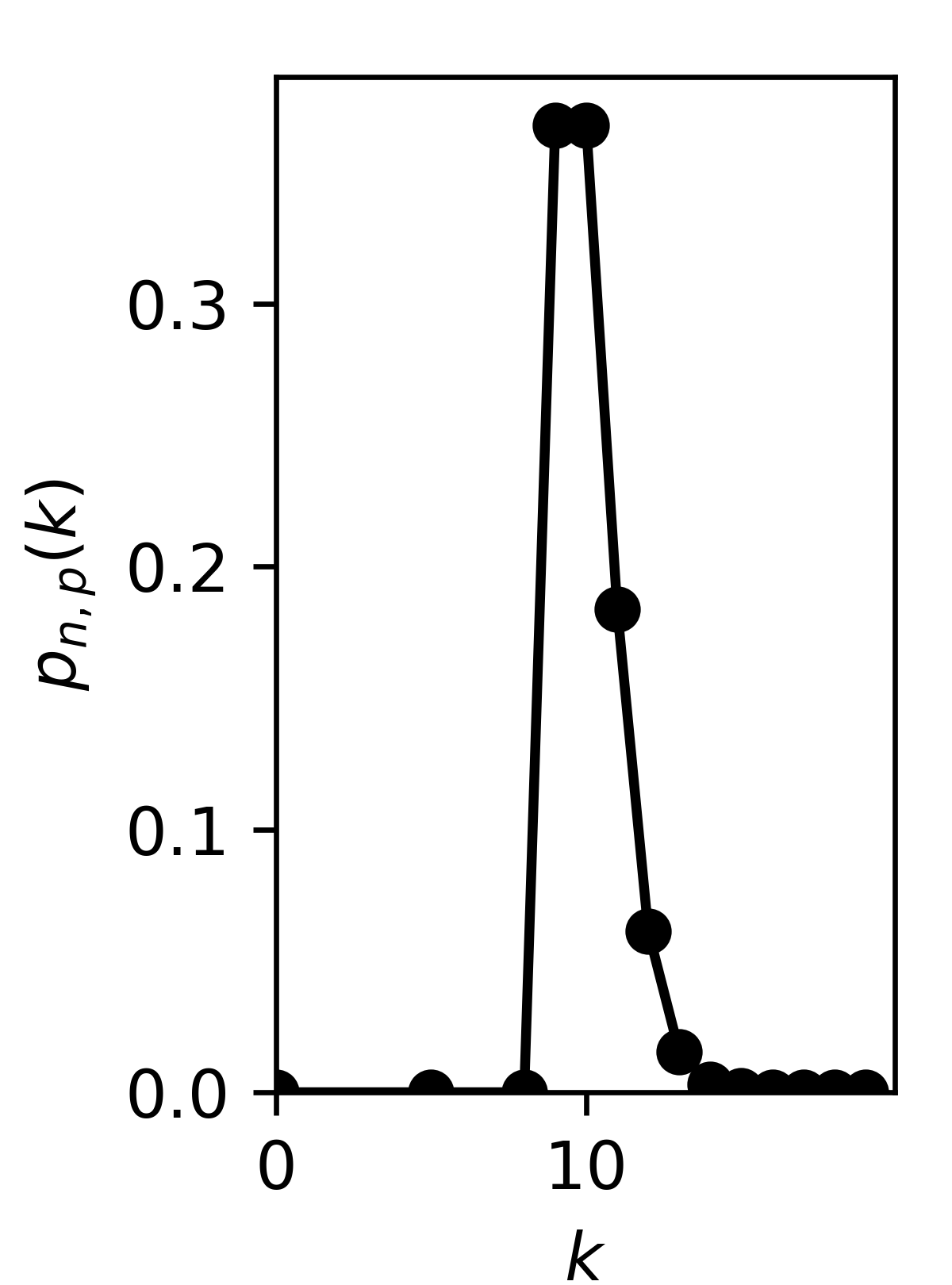}
\includegraphics[scale=0.95]{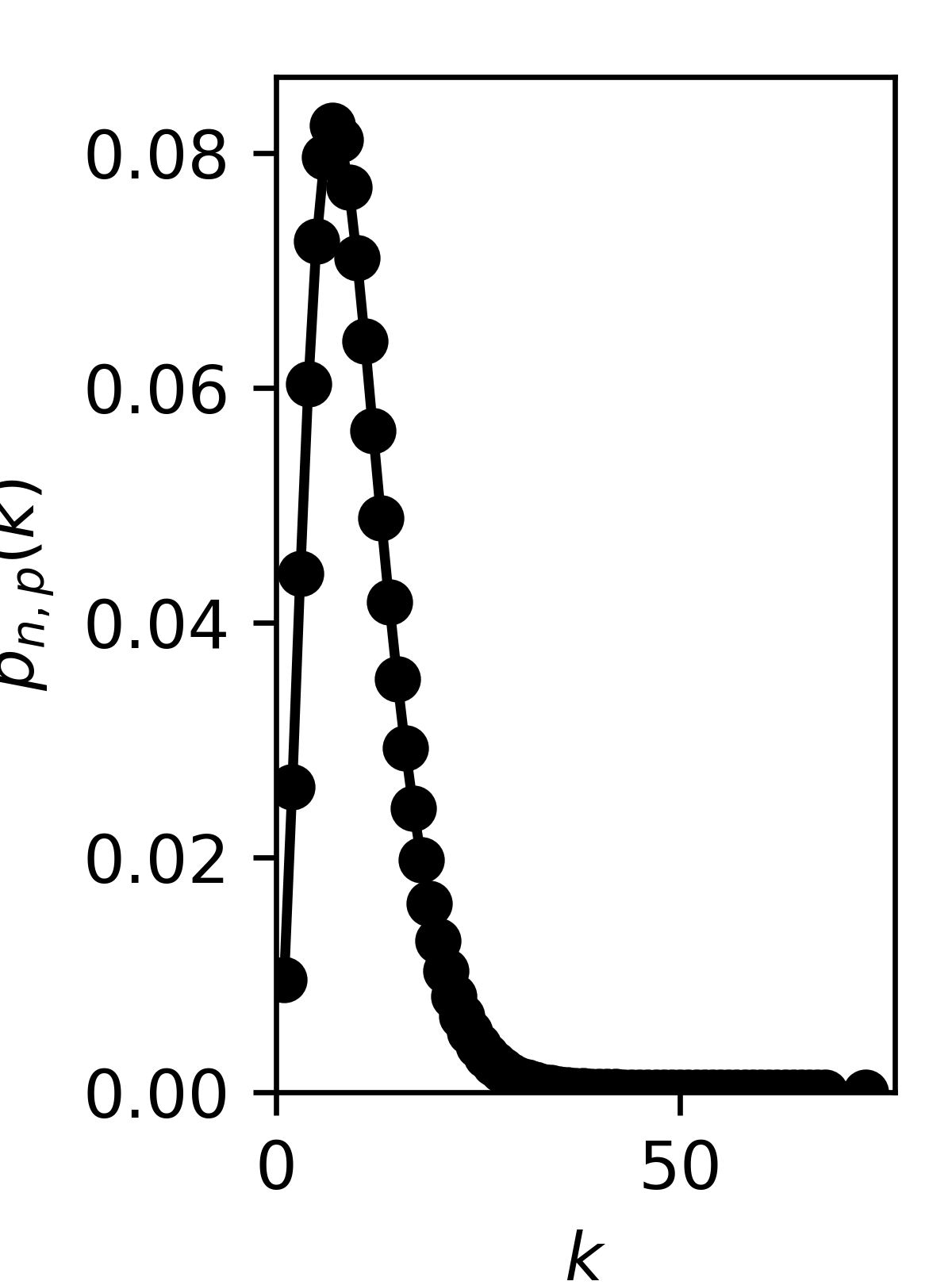}
\includegraphics[scale=0.95]{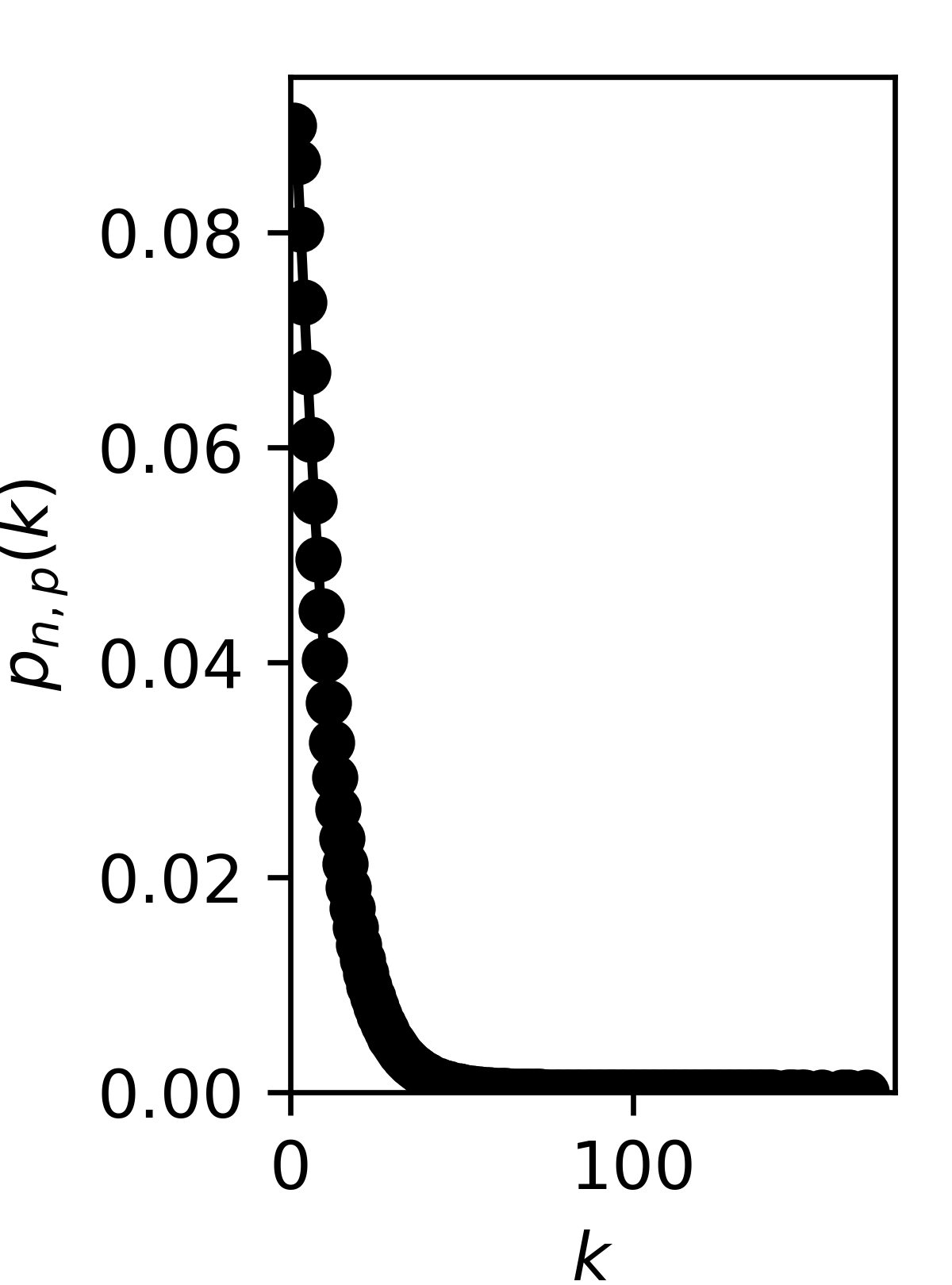}
\caption{The three degree distributions that constitute the benchmark.  $\langle k \rangle = 10$ in all the panels, where the variace is tuned by the parameter $p$ in ~\eqref{eq:neg-bin} to be, from left to right, $\sigma^2 = 1$, $\sigma^2 = 30$ and $\sigma^2 = 300$, respectively.} 
\label{fig:degreedist}
\end{figure}
\subsection{Model for disease induced herd immunity}
The main focus here is to investigate the impact of degree-heterogeneity and clustering on herd immunity induced by the first wave of the epidemic, also known as disease induced herd immunity (DIHI)~\cite{Britton2020}. In networks with heterogeneous degrees, the epidemic typically finds the high-risk groups first and thus `removes' important individuals or risk groups. In line with~\cite{Britton2020,gomes2020individual}, we exploit this fact and consider different levels of degree-heterogeneity using the degree-based mean-field and heterogeneous pairwise models to explore what happens in the wake of a lockdown period when some level of spreading is possible.
For illustrative purposes, we set the degree distribution of the network to be a negative binomial of the form
\begin{equation}
    P_{n,p}(k) = {k+n-1 \choose n-1} p^n (1-p)^k.
    \label{eq:neg-bin}
\end{equation}
The reason for this choice is that we want to highlight how heterogeneities in the contact structure play a central role in determining the DIHI. To illustrate this point, we typically consider three different degree distributions of increasing heterogeneity. An example of this is shown in Fig.~\ref{fig:degreedist} with $\langle k \rangle=n(1-p)/p = 10$, while the variance is tuned using the second free parameter. 

\begin{figure}
\includegraphics[scale=0.75]{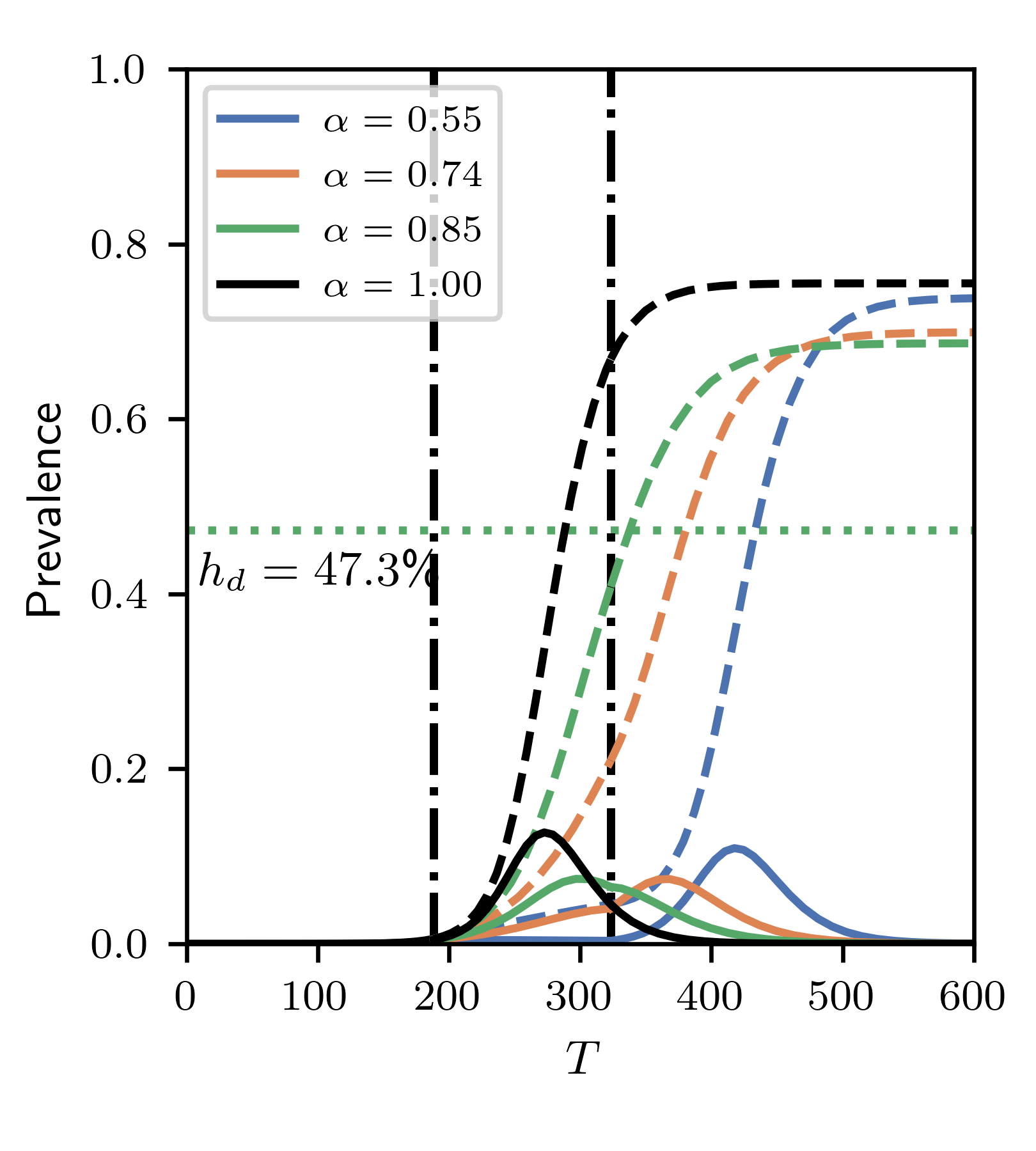}
\includegraphics[scale=0.75]{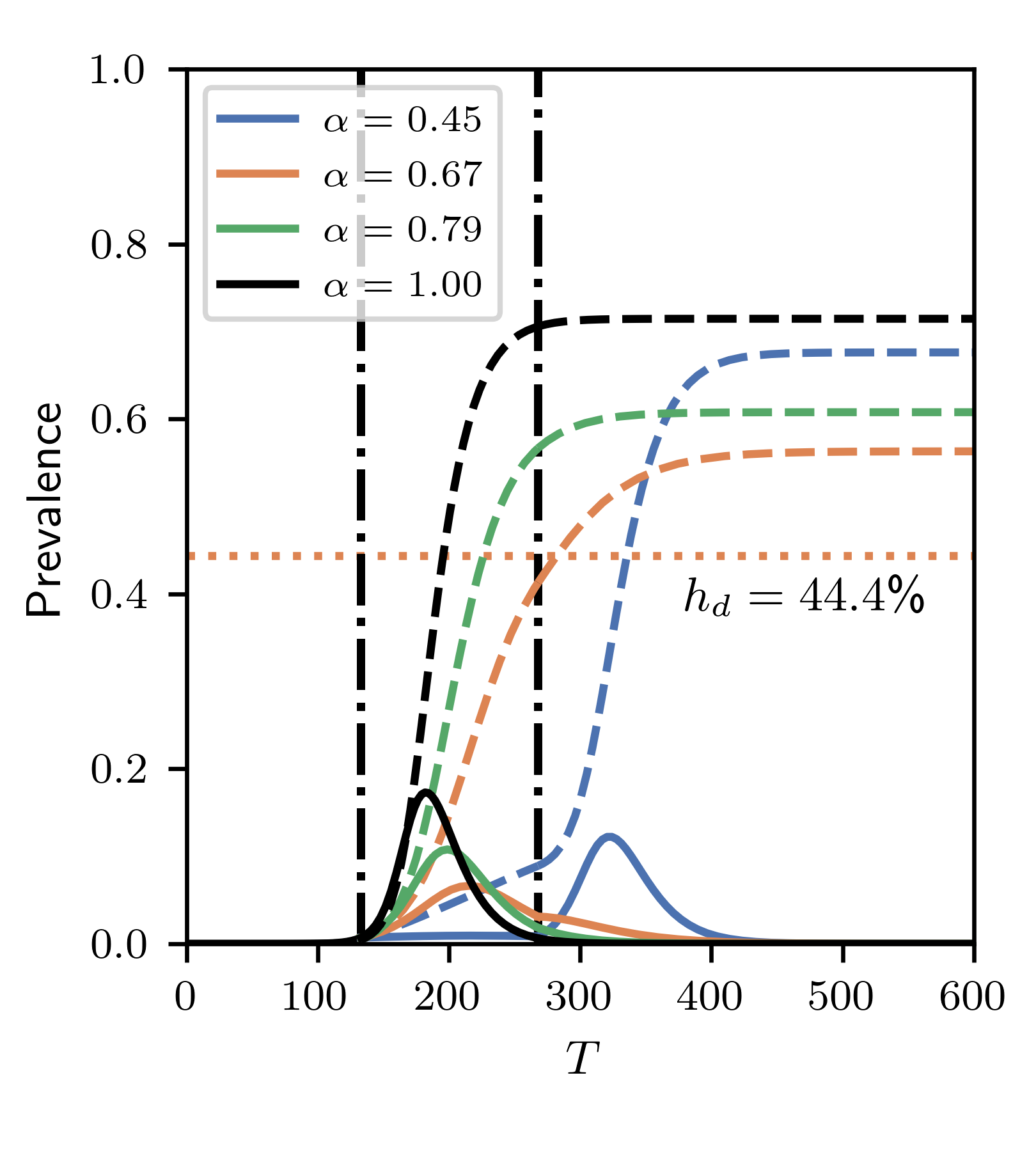}
\includegraphics[scale=0.75]{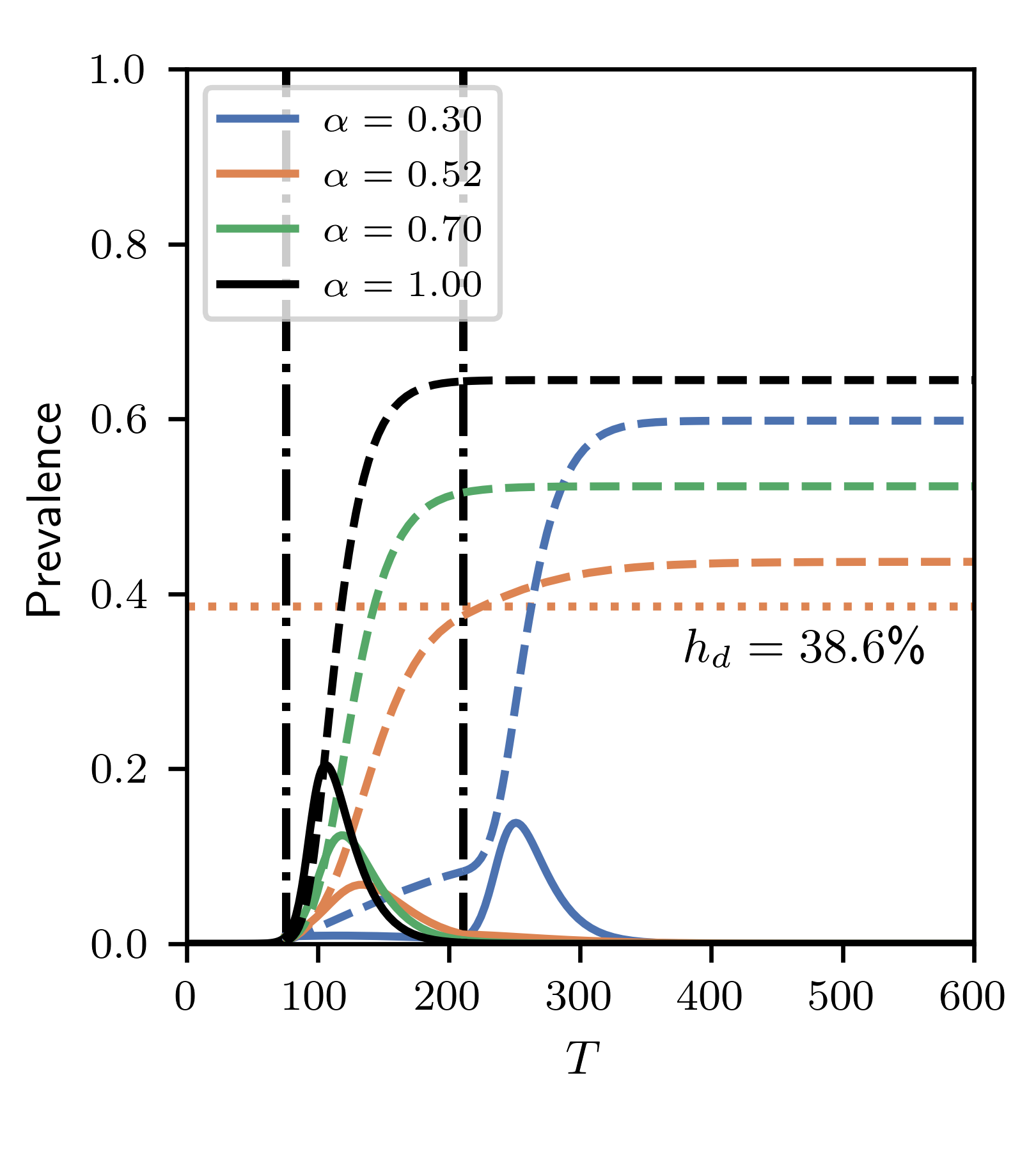}
\caption{Optimal $\alpha$ (see legend) and DIHI in delta-like (left), normal-like (centre) and scale-free-like (right) networks using the heterogeneous pairwise model with $\varphi=0$. Continuous curves indicate $[I](t)$, while dashed curves indicate $[R](t)$. The two vertical curves represent the beginning and the end of the lockdown. Horizontal lines and the corresponding percentages are the cumulative prevalence at the end of lockdown for the optimal strategy. Here, $\langle k \rangle=10$ and $\tau=0.03$.}
    \label{fig:epidemics}
\end{figure}

\begin{figure}
\centering
\includegraphics[scale=0.75]{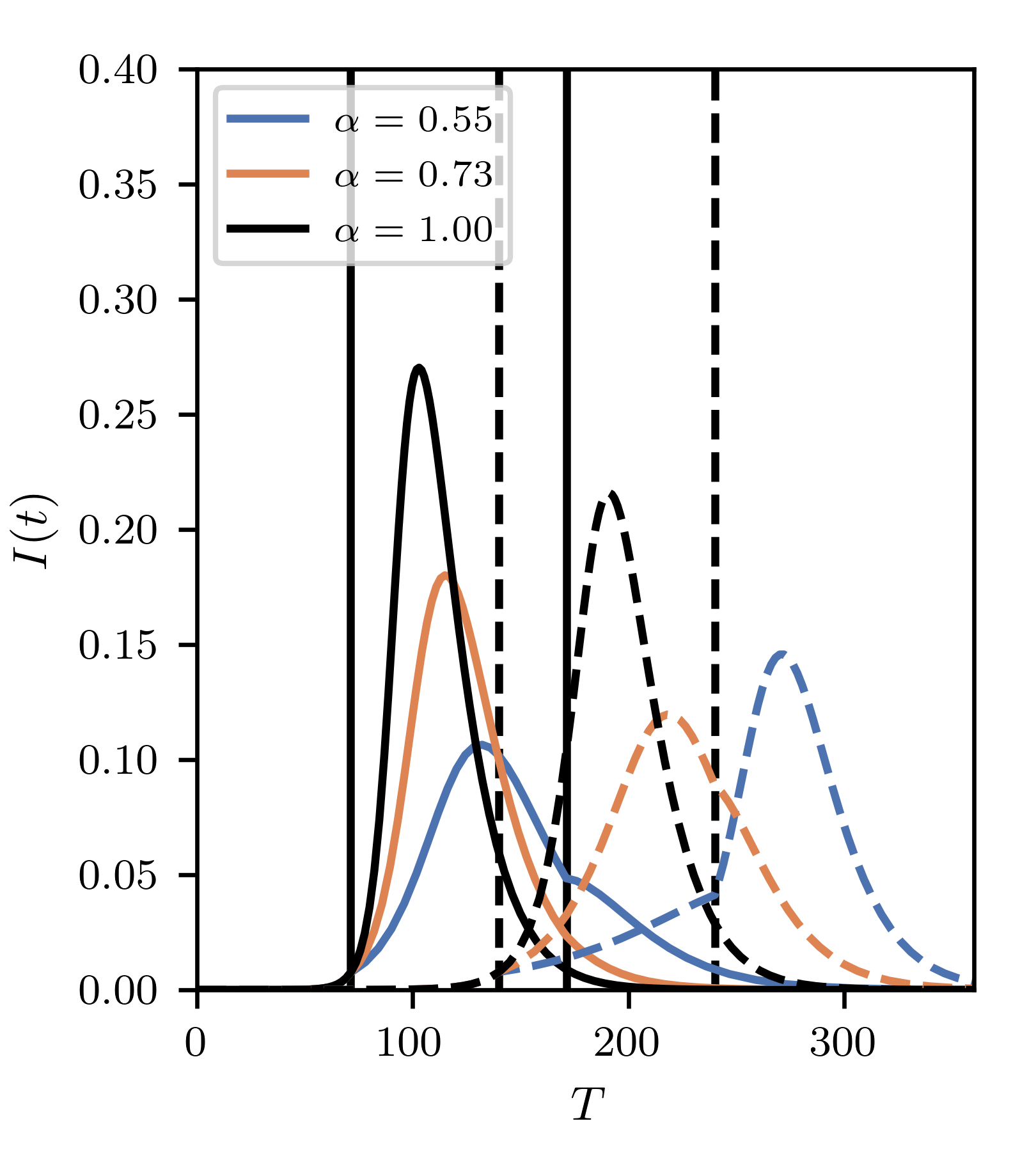}
\includegraphics[scale=0.75]{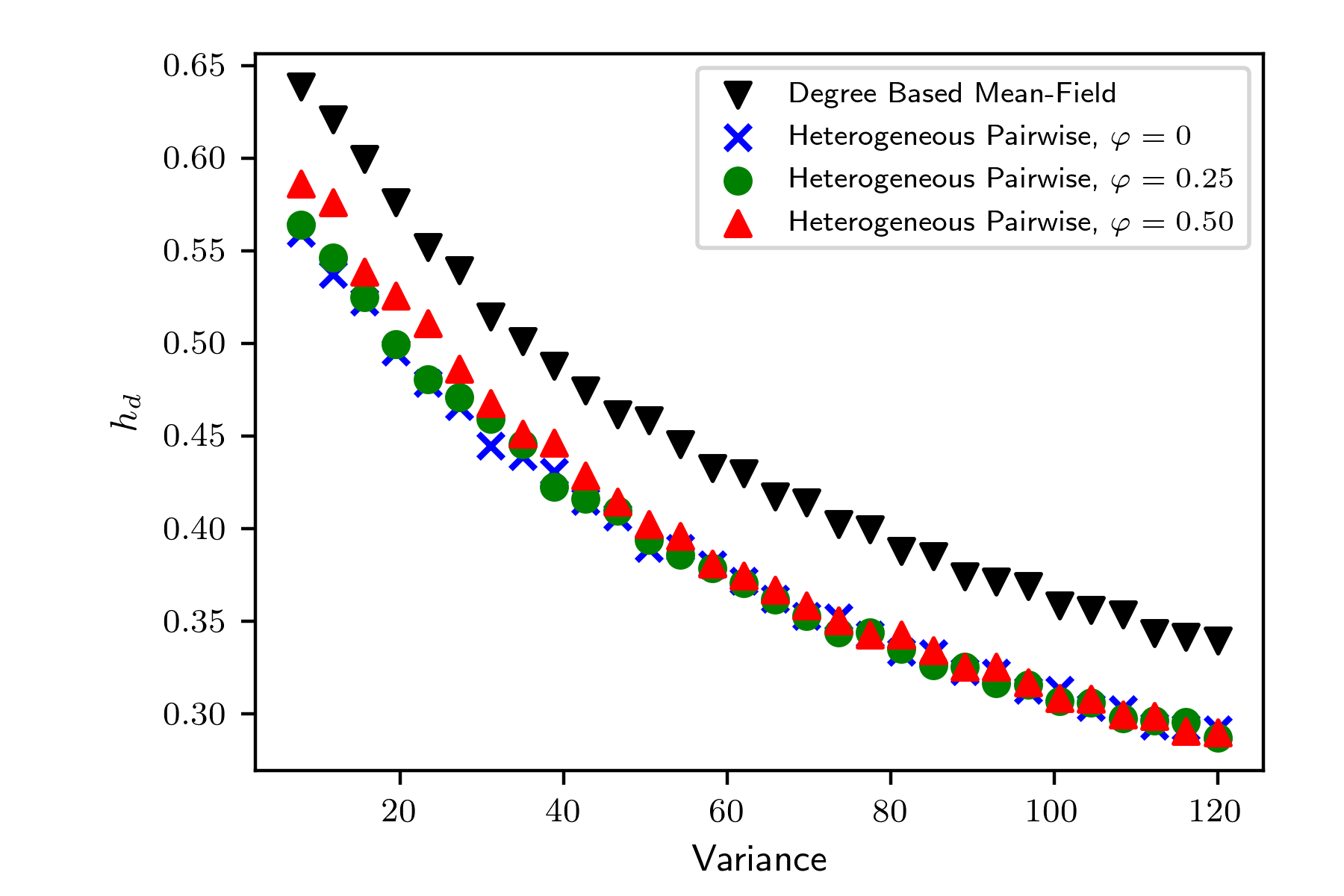}
\caption{(left) Difference between control acting on un-clustered networks (continuous lines) and clustered networks (dashed lines), with clustering coefficient $\varphi=0.5$.  Vertical lines are at the beginning (continuous) and end (dashed) of control. The blue continuous curve is optimal control for $\varphi=0$, the dashed brown is optimal control for $\varphi=0.5$. For comparison, the continuous brown is the optimal control for $\varphi = 0$ when applied to a network with $\varphi = 0.5$, the dashed blue line vice-versa. (right) Impact of variance in degree distribution on DIHI $h_d$, for different pairwise models with different values of $\varphi$ (see the legend). Control duration is 100 days from the moment $I(t)+R(t)\geq 0.025$. For both figures, average degree is $6$ and $\tau = 0.04$. The variance of the degree distribution used for the left panel is $15$, corresponding to the second point on the x-axis in the right panel.}
\label{fig:hetpwvariance}
\end{figure}
We consider a SIR outbreak on a fixed population of size $N=6.65\cdot 10^6$ (in line with many western countries, such as the UK). We arbitrarily set the recovery rate $\gamma = \frac{1}{14}$ and the per-contact rate of infection and average degree are given in the figure captions.  We initialize the outbreak by infecting $I_0=5$ nodes at random in one of the compartments and let the epidemic run until $0.5\%$ of the population gets infected. Then, a lockdown policy of duration $T$ reduces $\tau \to \tilde{\tau}_0 = \alpha \tau$. Afterwards, lockdown is lifted and $\tau$ returns immediately to its pre-lockdown value. 

Fig.~\ref{fig:epidemics} shows results based on the heterogeneous pairwise model without clustering for networks with increasing levels of degree heterogeneity (from left to right).
 Similar results (not shown) hold for the degree-based mean-field models~\cite{di2020impact}, with generally higher epidemics in the latter. In each case, we find the optimal $\alpha$ (a simple down scaling of the transmission rate without change to the network) and report the number of infections required to achieve DIHI  (i.e. total number of infected and recovered nodes at the end of lockdown such that the epidemic after lockdown is subcritical). Increasing degree heterogeneity leads to consistently lower DIHI levels meaning that fewer highly connected nodes play a more marked role in the transmission. Hence, loosely speaking as degree heterogeneity increases, it is enough to de-activate the most highly connected nodes in order to prevent a second wave.

It is interesting to notice that, aggressive control (low value of $\alpha$) leads always to a sustained second wave. Conversely, if lockdown is not strict enough (high value of $\alpha$) the epidemic will run its course during the first wave with some reduction in the final size. Hence, there is an optimal value of $\alpha$ for which the final epidemic size is smallest and the epidemic post-lockdown is subcritical.

If we include a clustering coefficient bigger than $0$ in the analysis, Fig.~\ref{fig:hetpwvariance}, we generally observe longer durations of the epidemic and smaller peak prevalence, compared to the unclustered case (see also ~\cite{volz:clustered_result}), with an overall smaller final size. This suggests that, all else being equal, it is possible to achieve the same herd immunity level with less aggressive lockdown measures, if the network is clustered.  It is worth noting that the final epidemic size is also smallest at the optimal $\alpha$ value (see also~\cite{Britton2020}).

Opting for the more accurate heterogeneous pairwise model, the level of DIHI is plotted for increasing values of variance in the degree distribution and for different clustering levels, see Fig.~\ref{fig:hetpwvariance}. It is clear that higher variance can drive DIHI levels to as low as $30\%$. The impact of clustering tends to lower the DIHI, but its effect is negated if the variance is high. This shows the non-trivial interactions between network properties where clustering has biggest impact in sparse networks and where high levels of degree heterogeneity can washout the effect of clustering. Furthermore, the same plot shows the DIHI levels based on the degree-based mean-field model. The trend is in line with results based on the heterogeneous pairwise model albeit with somewhat higher DIHI levels. However, the DIHI-levels are is sharp contrast, that is being much smaller, compared to $1-1/\mathcal{R}_0\simeq 0.76$ ($\mathcal{R}_0=\langle k \rangle \tau /\gamma \simeq 3.36$).

Finally, a note of remark goes to model selection. In the right panel of Fig.~\ref{fig:hetpwvariance} we see that the same epidemic on different network models (and across different mean-field approximations) lead to DIHI levels that range from a minimum of $30\%$ to a maximum $60\%$. This dramatic variation in the outcome of our analysis invites modelers to remain cautious when claiming that some results apply to real epidemics, as relying on approximating models might be misleading.

Several other observations can be made. First, heterogeneity here has been considered in contacts but similar effects will be obtained if heterogeneity in susceptibility is modeled or even if a continuum rather than a discrete range of heterogeneity is used. Second, contact heterogeneity is intrinsically linked to super-spreading but in case of COVID-19 the super spreading is linked to events/opportunities/context rather than individuals as it is the case for sexually transmitted infections. For example, COVID-19 super spreading between members of a choir or abattoir workers is more about the context and environment rather that the characteristics of an individual. Of course, some individuals are more likely than others to play a part in such events but it is more difficult to map out who the super spreaders may be. Thirdly, in reality during control/lockdown the network itself changes and this is not considered in the models presented here. Instead, here we kept the same network and reduced the rate of transmission across the same links. This can be very different from what happens in reality where the network changes due to school and workplace closures, for example. In~\cite{di2020impact} this has been considered and it was found that when the network during lockdown changed, DIHI levels were higher compared to the case when the network during lockdown was fixed. This was strongly dependent on the amount of community versus household links. Finally,  disease-induced herd immunity levels coming from various models are still high to be considered a viable way to control the course of the current pandemic and hence we should not be lulled into a false sense of security by such results and indeed we should all play a major part in observing and complying with any measure that limit transmission.

\section{Criticality and Network Effects on Epidemic Containment Measures}
\label{sec:bianconi}
\subsection{The challenge of modelling COVID-19}

The theoretical interpretation  of the data on the COVID-19 epidemics 
has proven to be very challenging.  The data quality, the testing policies and the methodology to record fatalities varies widely among different countries. These effects are very  significant  and  allow a  true comparison of the time series of infected individuals  only within a country. Among  different countries, despite the fact that comparisons based on infected individuals are  done routinely by news outlets, only the comparison based  on excess deaths data seem to provide an unbiased measure of the global impact of the epidemics in the society. 
Also if we neglect the challenges connected with the data quality modelling COVID-19 need to face important other factors. 
As the first pandemics in a global and connected human society, different factors play their role in determining the  efficiency of forecasting algorithms including containment measures, adaptive behavior of the populations and opinion dynamics.
Consequently, for  scientists working on epidemic spreading, predicting the evolution of the pandemic is a continuous effort of including data about human contacts and behavior into the model, inform the governments and the population, and then adapt again the model to the novel adaptive response of the population resulting in a ``weather forecast" of the epidemic spreading.
This type of research is quite suitable for Agent-Based-Models and Network Science models \cite{colizza2007reaction} at the meta-population level which include a compartmental description of the society and the mobility of the populations across different spatial regions.

Another challenge posed by the COVID-19 pandemics is that COVID-19 is a airborne disease. This  implies that the spreading routes are strongly affected by spatial proximity as encountered in urban settings. These include  public transport (underground, urban trains, buses) social activities (pubs, gyms, theatres, clubs, choruses), work places (major companies, universities, banks) healthcare spots (hospitals, surgeries, clinics).

Spatial proximity has been investigated in experiments recording face-to-face contacts such as in SocioPattern experiments \cite{isella2011s,stehle2011high,vanhems2013estimating}, however typically  the contacts modelled by network science have focused on interactions characterizing social ties, such as  friendships or acquaintances. 
Consequently,  most of the   realistic attempts to describe the pathways of the  epidemic spreading  of COVID-19 consider coarse grained and aggregated information rather than existing models of social networks formulated to capture social ties.
The most relevant exception to this rule is the modelling and the treatment of automated tracing data and the resulting tracking of the epidemics, which provide a solid benchmark for modelling frameworks \cite{cencetti2020digital}.

In the following paragraph we highlight important theoretical insights that can help clarify important aspects of the models used.  Specifically  we touch on the problem of modelling epidemic plateaux and of using network theory for predicting the efficiency of track and tracing apps.

\subsection{Containment measures, plateauing time-series and  criticality}

\begin{figure}[ht]
	\begin{center}
 \includegraphics[width=0.95\columnwidth]{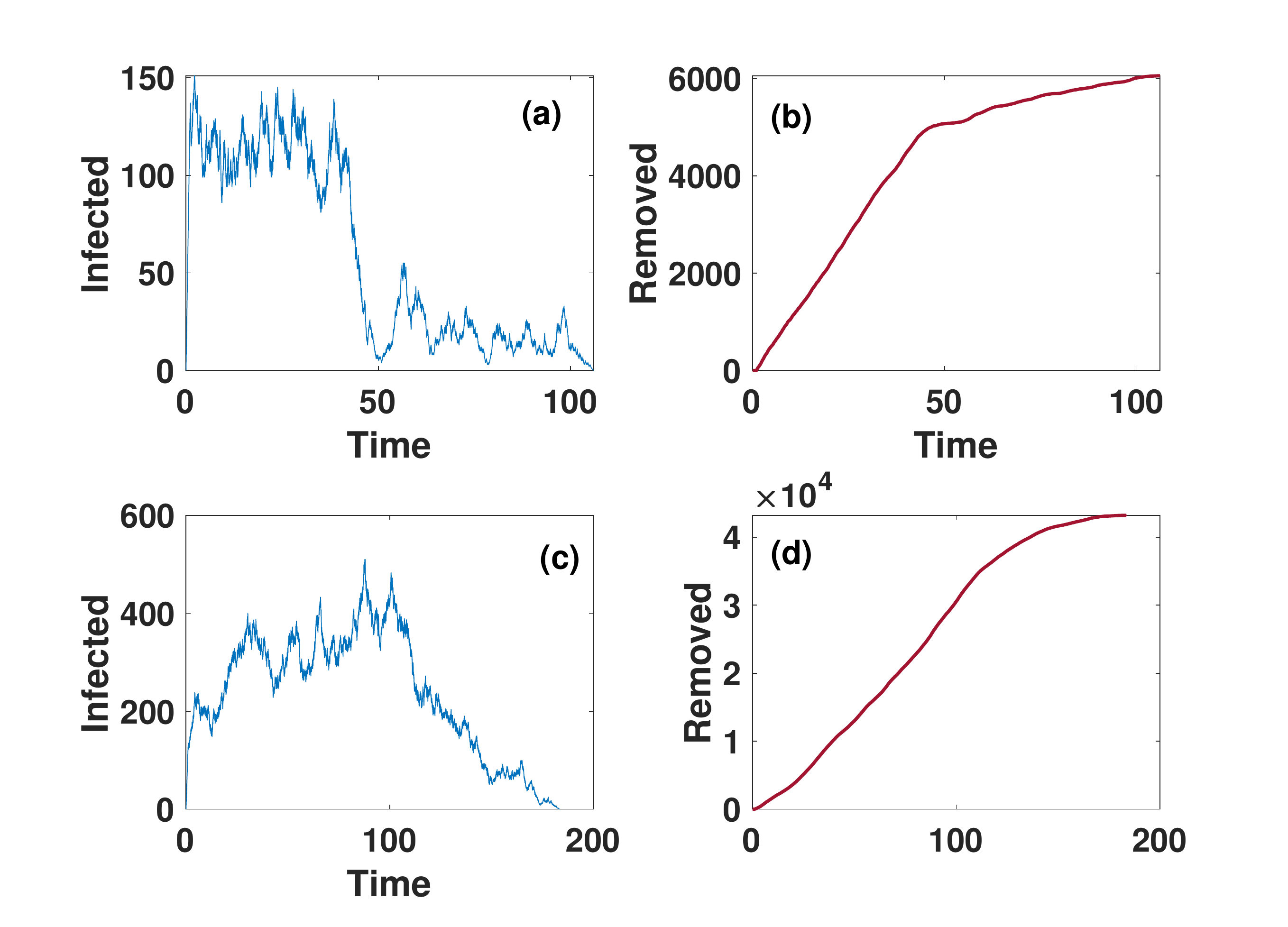}	
  \end{center}
  \caption{Epidemic time series generated by the stochastic SIR model at criticality starting from  non-trivial initial conditions. Panel (a,c) show two  time series of the number of infected individuals  while panels (b,d)  show the corresponding time series of removed individuals. All time series correspond to a population of $N=10^6$ individuals and an initial number of infected individuals given by $n_0=128$.Despite the panels (a,b) and (c,d) are generated using the same model with the same parameters the resulting two SIR dynamics are significantly different. In particular   the outbreak size and the outbreak duration are very different in the two simulations due to stochastic effects. }
  	\label{fig:plateau}
      \end{figure}

At the onset of the COVID-19 pandemics the time series of infected individuals and deaths clearly followed exponential growth at each epidemic focus.
The doubling time of these exponentials ranged between two and three days in Europe at the onset of the epidemics providing evidence for a similar spreading dynamics well captured  by the SIR dynamics in well-mixed populations. 
To "flatten the curve" of the number of infected individuals two types of containment measures were  adopted. The first one, includes the lock down and  focuses on reducing the number of contacts of each individuals. The second one implies a fast detection of cases and includes efficient track and tracing strategies \cite{ferretti,bianconi2020epidemics}.
After the first stages of the evolution of the epidemics, when containment measures have been implemented, epidemic time series started to show characteristic plateaux not typically encountered in epidemic models \cite{bianconi2020epidemics}.
The typical SIR evolution of an epidemics with constant infectivity $R_0$ includes an exponential onset of the number of infected individuals, and an epidemic peak marking the characteristic time at which the infection has spread to a large fraction of the population, producing herd immunity and causing the reduction of the number of infected individuals in time. This is the scenario expected in the SIR dynamics when we are in the so called {\em supecritical regime} with  infectivity $R_0>1$. In this case the epidemic spreading result in  the infection of a finite fraction of the population, leading to many fatalities if containment measures are not put in place to control the spread of the virus. A particular feature of the SIR time series in this supercritical regime is that the peak is well defined and not plateauing as long as $R_0>1$.
In Ref. \cite{radicchi2020epidemic} an explicit calculation shows that plateauing time-series are generated only if the epidemics is  nearly {\em critical} with $R_0\simeq 1$ and the population is far from herd immunity. 
The critical regime of epidemic spreading has been investigated in statistical physics and mathematical biology starting from a single initial seed \cite{zapperi1995self,ben2004size}. In the context of COVID-19 the critical regime however Ref.
\cite{radicchi2020epidemic} points out that the critical regime can be reached dynamically with containment measures at the later stage of the epidemics when the number $n_0$ of infected individuals is greater than one, i.e. $n_0>1$. This critical SIR dynamics is characterized by a power-law growth of the number of removed individuals reported in several countries at the later stage of the epidemics  \cite{Ziff,Blasius} and is  strongly affected by fluctuations,  which make predictions of the duration of the outbreak and their size very challenging. For this reason in this regime  it is crucial to abandon deterministic modelling of epidemics  and embrace a full stochastic modelling of the epidemic spread \cite{radicchi2020epidemic}.
In Fig. \ref{fig:plateau} we show two examples of stochastic time series of the SIR critical dynamics showing the important effect of stochasticity and providing evidence that plateauing time series can spontaneously occur for critical SIR dynamics with non-trivial initial condition.

In order to describe the observed COVID-19 plateauing time series many modellers presently consider adaptive models that consist in increasing and decreasing the infectivity $R_0$ in time. It is possible that human adaptive behavior can be modelled in this way, however this is not a necessary assumption to obtain plateauing time series. Moreover most of these ad hoc models might still strongly underestimating the role of fluctuations as long as they rely on deterministic models.

\begin{figure}[ht]
	\begin{center}
 \includegraphics[width=0.95\columnwidth]{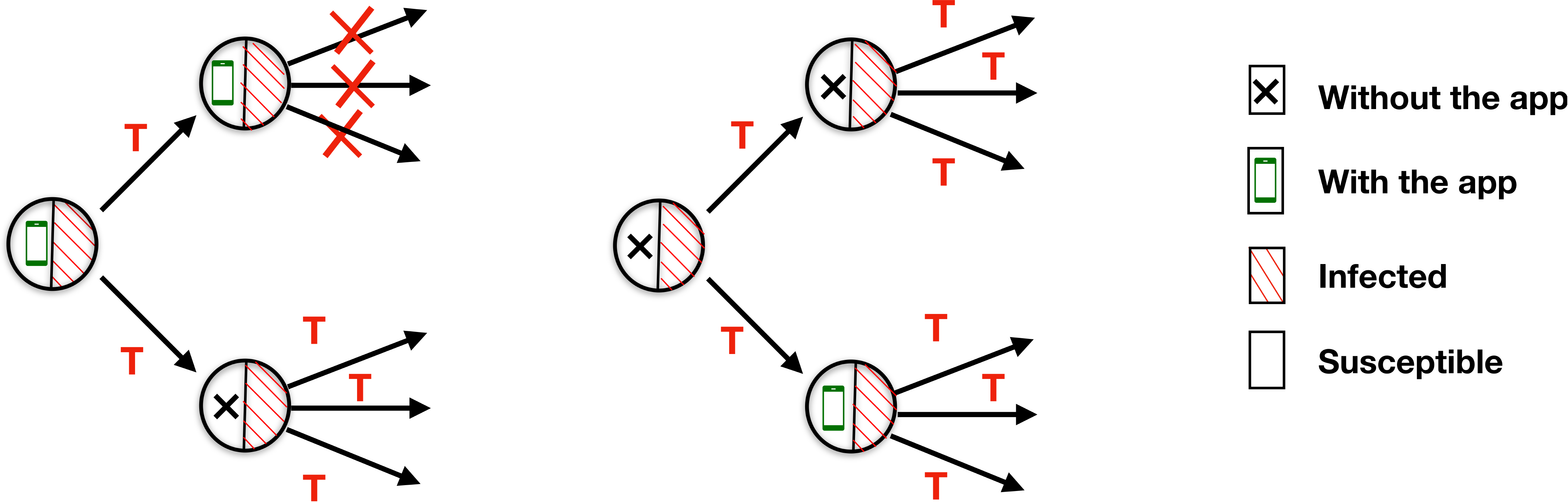}	
  \end{center}
  \caption{Automated tracking and tracing can be modelled by mapping the SIR to percolation capturing the non-linear effects of the spreading dynamics. In this model each  individual is assigned two states: the first label indicate if the individual is infected or susceptible, the second label indicates if the individual has or does not have the tracing app. The epidemics spreads from each infected individual to a neighbour susceptible individual with probability $T$ called the {\em transmissibility} unless the infected individual has the app and has been infected by another individual with the app as discussed  in Ref. \cite{bianconi2020message}.}
  	\label{fig:diagram}
      \end{figure}

\subsection{The network effects in automated track and trace}

Determining what is the  fraction of adoption of tracing apps that would guarantee a good efficiency of the technology and the control of the epidemic spreading  is a fundamental problem in COVID-19 research activity.

Despite the problem is a inherently network problem most of the attempts to address this problem rely exclusively on linear dynamics \cite{Ferrettieabb6936,Fraser6146}.

Network theory can provide  an important contribution by allowing to  capture non-linear effects of the epidemic spreading thanks to the mapping of this process to percolation \cite{Newman_messagepass_pre_2010,karrer2014percolation,
bianconi2018multilayer}.

In a recent paper \cite{bianconi2020message} the mathematical framework to fully capture the role that automated tracing app have on epidemic spreading has been proposed.
In this model every node/individual  of the network is assigned a variable  indicating whether the node has adopted or not the app.
 Infected individuals  transmit the disease to a  susceptible neighbours with probability $T$, called the {\em transmissibility} of the epidemic unless they have the app and they have been   infected by  individuals with the app (see Fig. \ref{fig:diagram}).  This theoretical model fully capture all the non-linear nature of the spreading process and can be solved on real networks using message passing techniques.
This model indicates that also a moderate adoption of the tracing app can have a significant impact in slowing down the spread of the epidemics \cite{bianconi2020message}. 

\section{Agent-based modeling}
\label{sec:prokopenko}

Stochastic agent-based models (ABM) have been successfully used for modeling the COVID-19 pandemic, evaluating non-pharmaceutical intervention strategies, as well as providing timely policy advice~\cite{Ferg2020,Chang2020}. For example, the ABM approach to tracing and controlling the pandemic in Australia~\cite{Chang2020} compared several mitigation and suppression strategies, and pinpointed an actionable transition across the levels of social distancing compliance, in the range between 70\% and 80\% levels. Specifically, a compliance at any level below 70\% was shown to be insufficient to reduce incidence and prevalence, for any duration of social distancing, even when coupled with effective mitigation (i.e., case isolation, home quarantine), and international travel restrictions. In contrast, under the same mitigation and border control conditions, a compliance at the 90\% level was found to control the disease within 13--14 weeks. In addition, this ABM accurately predicted key features of the first wave in Australia: the peaks of incidence and prevalence in late March and early April 2020 respectively, and the range of cumulative incidence attained after the suppression period at the end of June 2020. It also identified formation of the second wave in early July 2020, once the suppression strategy is relaxed~\cite{Chang2020} .  

This predictive accuracy was achieved by utilizing a high-resolution individual-based computational model calibrated to both demographic features of the Australian population (based on the Australian census data) and key characteristics of the COVID-19 pandemic. The demographic component was validated previously, in context of pandemic influenza modeling~\cite{Cliff2018,Zachreson2018,Harding2020,Zachreson2020}, while the COVID-19 epidemiological component was validated in a now-casting mode in March 2020. Furthermore, the model was cross-validated by a genomic analysis of COVID-19 activity in New South Wales (NSW), the most populous state of Australia~\cite{rockett2020revealing}, focused on locally acquired clusters in the state. In particular, the fractions of local transmissions inferred by the ABM were compared against the genomic sequencing of SARS-CoV-2, carried out during February--March 2020 in a subpopulation of infected patients. During this period, the confirmed cases in NSW comprised 44.5\% of all cases detected nationally. Only a quarter of sequenced cases were identified as locally acquired. This was in a good agreement with the traces obtained from the ABM: the transmissions across households and household clusters amounted to 18.6\% (std. dev. 2.9\%), and the upper bound including, in addition, all the transmissions across local government areas was estimated at 34.9\% (std. dev. 8.2\%)~\cite{rockett2020revealing}.

The capacity of ABMs to represent commuting patterns and interactions within and across multiple mixing contexts, ranging from households to local areas to workplaces, is one of their strengths. This quality is shared by ABMs and network-based approaches to epidemic modeling~\cite{newman2002spread,cauchemez2011role}, as both types of models attempt to capture fine-grained interactions and contextualized transmission contexts --- unlike canonical compartmental models which essentially assume a fully mixed contact topology. Nevertheless, these models describe the context dependence and interventions differently. In ABMs, heterogeneous social mixing is modeled by explicitly specifying the categories, e.g., households, workplaces, schools, etc., following a relevant pandemic maxim: ``same storm, different boats''. Consequently, various social distancing interventions are described with the corresponding macro- and micro-distancing parameters, e.g., 80\% of agents comply with social distancing by reducing the intensity of their interactions (and hence, the transmission probability) by 100\% at the workplace and by 50\% within the local community. In network-based models, each edge typically represents a specific transmission route, and the interventions are modeled by topological changes, aiming to reduce the diversity of interactions~\cite{meyers2003applying,small2020modelling}. Thus, both model classes enable analysis of both time- and context-dependent interventions, which may also include counter-factual and hypothetical scenarios. 

The high-resolution spatial and temporal representations adopted by ABMs and network-based models offer another strength: the capacity to comprehensively examine the space of ``control'' parameters and identify the corresponding phase transitions (tipping points)~\cite{yeomans1992statistical}. For example, critical regimes have been previously identified in epidemic models, which interpreted an epidemic spread as percolation through a complex network~\cite{newman1999scaling,newman2002spread,balcan2011phase,Harding2018,Harding2020b}. Similarly, as pointed out by Chang et al.~\cite{Chang2020}, the transition across the levels of social distancing compliance can be seen as an example of a percolation transition in a forest-fire model with immune trees~\cite{Guisoni2011}. 

The benefits of high-resolution modeling and comprehensive exploration of the relevant phase-spaces come at a price: the relatively high computational burden of ABMs relying on detailed datasets, the need to carefully calibrate the numerous ABM input parameters across various mixing contexts, or the necessity to specify detailed network interactions, while varying network topologies. Nevertheless, most of these drawbacks can be addressed by an ever-increasing computational power of high-performance computing and modern data science. In addition, some of the actionable results, e.g., phase transitions, may be obtained in simplified settings which concisely capture the most salient features of the modeling problems.

There are several practical modeling questions that present immediate challenges for both ABMs and network models. Most of these are related to difficulties in accounting for the human factor: how to infer friendship networks which affect social interactions; how to account for human mobility, including long-distance travel, under various local lockdown regimes; what are the best ways to estimate in-hospital and in-quarantine transmissions; how to realistically model contact tracing under capacity constraints, and so on. 

However, an emerging understanding shared by the COVID-19 modeling community is that there is a pressing need to start developing ``pro-active'', rather than ``reactive'', models. In other words, the COVID-19 modelers must aim at the issues that the society is likely to face in the near- to mid-term. The much anticipated variety of new vaccines and treatments brings the need to optimize, while satisfying complex immunization priorities, over complex regimes. These integrative regimes may combine several pharmaceutical interventions, such as vaccination and targeted antiviral prophylaxis~\cite{longini2005containing,germann2006mitigation,Zachreson2020}. Some intervention policies may innovate with ``interaction substitution'', e.g., ``shield immunity'', in which recovered individuals are deployed at focal points of essential services~\cite{weitz2020modeling}, or with controlled human infection trials~\cite{shah2020ethics}, bringing about logistical, social and ethical challenges. The novel coronavirus may develop genetic variations, increasing the risk of endemic transmissions of SARS-CoV-2 and its variants, which in turn may interact with existing coronaviruses and emerging zoonotic diseases~\cite{kissler2020projecting,salata2019coronaviruses}. Concurrently with health risks, the diverse regional and international impacts of the pandemic create complex dynamics of reconfiguring trade networks and emergent travel bubbles~\cite{Lenzen2020,lee2020global}. All these challenges will continue to unfold in presence of worsening systemic socioeconomic vulnerabilities and overstressed healthcare systems. To address these questions, we need to close a significant gap in the understanding of systemic risks in complex social systems with nonlinear dynamics, while focusing on identifying early precursors for tipping points and critical transitions~\cite{scheffer2009early}. 

In summary, no single modeling approach will provide all the answers, nor is there a single clever exit strategy. Our models and crisis management frameworks must start adapting faster than the virus, offering pro-active and ``smart'' interventions. This should be complemented by tailored disease surveillance and public communication policies, generating broad social engagement with the approaches.

\section{Overview of Contact Tracing Apps}
\label{sec:crowcroft}
\subsection{Introduction}

Identity services ~\cite{dumortier2017regulation} are used to uniquely distinguish a specific person from others (“who are you?”).  
This allows the association of that unique person with attributes, establishing credentials to accessother services (“what is it to me?”). 
In this way such services afford entitled rights to individuals.

Digital identity systems replace, or enhance, traditional paper documents (such as passports, identity cards) with largely online computer based solutions.  
There can be a separation of the foundational system that provides the first step, from a functional identity system, that serves a specific application or purpose (e.g. voting, tax, health, driving, age verification etc etc), or they could both be combined.
Such digital identity systems can potentially make rights more readily and affordably available to more people. 

We elaborate here on one aspect of the implementation of digital identity systems and Covid apps, namely where do we choose to place the system,  and why we might make different choices of that design decision. Two  key considerations are resilience and security. Since identity underpins so many other systems in society, it behoves us to be clear about those considerations and the assumptions about trustworthiness that lie behind them.

The principle choices involve where we find key identity data. 
Both functional and foundational identity systems can be implemented in a number of ways.  
Two of the key categories of systems are centralized and decentralized

\begin{description}
\item[Decentralized Identity Systems ~\cite{german_eID_whitepaper} and decentralized  contact tracing~\cite{bell2020tracesecure}.]

Each person creates and curates their own data. There is no external service (deemed self sovereign). If someone wants to found out who I am and what I can do, they ask me.
Data still needs to be somehow ratified by some ‘authority’ in the first instance, but from then on you are your own authority. You can vouch for yourself.

\item[Centralized Identity Systems ~\cite{Aadhaar_2017} and centralized contact tracing~\cite{briers2020risk}.]

Everyone places his or her data in a server; colloquially sometimes referred to as “putting all your eggs in one basket”. 
A weakness in this is that if you drop the basket, or someone takes the basket, you have no eggs.

\end{description}

For centralized systems, not just for identity services,  in general we usually combine a number of technologies to provide assurances against certain problems such as the loss of confidentiality. 
For example, services such as Authentication, Access control, Authorization support,  and we hope encrypt the data at rest, in transit, and during processing. 
Modern systems can also run servers in Secure Enclaves, e.g. using Intel's SGX or ARM's Trustzone or similar, which provide for hardware enhancements to improve resistance against attacks on privacy. One can also employ software techniques such as Fully Homomorphic Encryption (FHE) to carry out the lookups, without decrypting the query or response or data being queried.  
In addition, Differential Privacy techniques permit you to control how much is revealed  by a query to a central system.
in some cases, for example when the querier wants to learn aggregate statistics, access is permitted, but is denied for a specific identity/attribute pair which can reveal personal data, (e.g. ``this person is over 21'').

Centralized systems may serve entire nations (e.g. for passports, birth certificates etc).
To provide credentials (foundational id and functions like visas for travel),  you need to federate these nationally centralized systems.  Federation takes a set of two or more national  services, and turns them into a distributed service.  You might expect this type of federation to offer less functional services than each of its component,  national system. Of course, decentralized systems also need to be federated at scale to be useful - after all, no one is an island.

For distributed identity systems different users' foundational identity is created and stored in different servers.  
Different users' functional identity services may be served from different servers. 
You can also use secure multiparty computations (MPC) to look up identity data that is split across server sites.

Of course distributed identity systems can also be federated. 
If you seek to identify groups of people, or combine multiple functional attributes for one or more individuals,  you will need to access multiple servers.
Of course you can also have hybrid identity systems where you do not keep all the data in one place (known as data minimization),  but rather federate/shard/distribute/ and decentralize the service.

Distribution, in turn, adds complexity. 
Fully centralized and decentralized systems will necessarily be simpler. 
Centralized and distributed  systems can offer higher availability, by virtue of replication and load balancing. 
Replication of data services for slow changing data is, in fact,  very simple. 
Hybrid systems may combine decentralized systems with central or distributed purely for the purpose of increasing availability so that if a user’s device is unreachable, broken or stolen, they are not deprived of identity services. 
Replicated services  may also better resist denial of service attacks.

Some of the debate around the choice between these approaches stems from security concerns. As outlined above, some simple attacks can be mitigated by a range of privacy enhancing technologies. Certain threats may be inherently easier to resist with some implementation patterns than with others.

Apps may or may not have identity needs. In some cases, we can replace actual, foundational, identity with tokens that act as randomized proxies for id. 

\begin{description}
\item[Symptom reporting]

No need for id, but needs uniqueness of reports - inherently it is about medical  stats at a given time ...token are more than good enough. There's never any need to re-link a report to an individual.

\item[Contact tracing]

Centralized and may require id, as the reason for centralized systems may be to combine health status with other factors (age, gender, ethnicity) for epidemiological studies and then relink to other ids to see if if there is any variation in infections between different groups. 
Indeed, one can also uncover immunity expiry (e.g. via re-infection, if and when that occurs).

Decentralized apps often only have a token as their design goal  to enforce the anonymity between an index case and exposed individuals (who may be unknown to the index case, e.g. someone who happened to be on same bus or in same bar).
Similarly this anonymity between exposed people and the health service provider, and other stakeholders, is desirable to prevent the abuse of compliance rules for observing self-isolation.  
An unfortunate consequence of this is that there is no store of data  to carry out epidemiological  statistics, as asymptom reporting app can.

\item[Immunity passporting]

May need id, if required e.g. as a visa to accompany a real passport to allow for travel: 
can be centralized, or could be  decentralized  where each user holds immunity status and just has to show it associated with a foundational id to verify.

\end{description}

\subsection{Design choice - driven by threats?}

Threats exist to the correct and trustworthy operations of Identity Systems (and apps). 
What are the threats, and from whom do they originate? 
This is not an exhaustive list, but to illustrate the range of considerations that might impact on the cost of various implementation patterns for id or a health app.

\begin{description}
\item[The Human level]
\begin{itemize}
\item
Fool the system (e.g that I am under 21, or  I am immune).
\item
Fool or coerce people to register/deregister ( commonly known as a masquerade), or require immunity passport to return to work, e.g. avoiding cost of  providing a  safe workplace..
\item
Fool or coerce  people to verify credentials on behalf of someone, i.e. spoofing.
\item
Run spoof service, so people give biometric, and other info, to a fake interface.
\end{itemize}

\item[The Technology level]

\begin{itemize}
\item
Does the system actually provide minimal answers (e.g. ``is over 21” not “is 24”)? 
Can the user have confidence that the system only displays an anodyne yes or no, and only the client knows what was queried (for example ``is your age over 21”)?
\item
Exploit vulnerabilities in service to do the above individually
\item
Id theft, masquerade, prevention of service, etc
\item
DDoS service across a wide range of servers...
\item
Indirect attacks (e.g. on network \& power infrastructures)
\end{itemize}

\item[The Organizational level (insider attacks, state actors etc)]
\begin{itemize}
\item
Surveillance of use (and meta data use e.g. what id is used when \& where)
\item
Surveillance of register/deregister (set membership attacks)
\item
Isolation of subgroups by attribute, for differential treatment...
\end{itemize}

\end{description}

What would make `id-as-service' more trustworthy?
Consider the following.
A client presents a key, and gets one or more values back. 
An example key is a biometric (e.g. pass phrase, fingerprint, iris, face, palm, etc) plus a possible additional parameter (e.g. age verify, bank account number). 
The response value is returned:  ``is over 21'', bank a/c, ``is entitled to NHS care'' etc. 
In some system designs, what is returned is a token (or collection of tokens) that have a sole purpose of authentication and are of no use or meaning to anyone else.
The client side should run with security, up to and possibly including client user context ,such as knowing who can see the display or know the location etc.
The network should  at least implement basic security such as Transport Layer Security (TLS). 
The server side should ensure and  keep all data encrypted.
It is possible to run the server in an enclave (SGX, Trustzone etc). 
However, a problem that can occur is if these are subsequently compromised, but we  continue to use anyway  (i.e. confidential cloud) with relatively low performance penalty.
Enclaves also potentially provide attestation (e.g. of integrity), which can also be useful but might depend on a single authority that has to be trusted and trustworthy.
We could run the server key/value lookup using FHE. 
In this case the problem is performance, look up rate could have a pretty low throughput. 
However, see this service which claims otherwise:- http://privatebiometrics.com/index.html .

We could run the server with data sliced or sharded (that is disaggregated), and use MPC to do match key to value.
This has some latency challenges, but is not computationally pathological, and scales out well.
Note others have built solutions to privacy in this space too, e.g. https://cryptpad.fr/

We could distribute data over many cloud services and federate.
Alternatively, it is possible to run a fully distributed bespoke system (possibly non virtualized/not cloud.  
The simplest would be to put the key/value store on a P2P Chord/Distributed Hash Table like Kademlia. 
This basically mimics regular password file (hash onto a file, but in this case, has onto a node in Kademlia). Kademlia also supports resilience/node failure recovery and has high performance. 
There is no access control inherently, but it could be added.

Another candidate for this is a distributed ledger system (DLT), such as Ethereum or Hyperledger, with one added benefit that this has high integrity, and is effectively tamper proof.
Ledgers can be fully peer to peer (p2p), and therefore permission-less, or depend upon an access control system that itself could be distributed or centralized (permissioned).  
Mutable data has to be kept off chain, or some new construct applied.
 DLT also support computation, as part of transactions, and IBM hase proposed adding MPC  as part of these computations.

There is a slight circularity here, in that  the sign on  for a permissioned system itself requires authorization.  
So, if the permissioned blockchain is supporting id-as-a-service, who provides the id for the sign-on? 
Note the entities using the service are as likely to need Id-as-a-service as the subjects  (the bank manager is a person too). 
Permissioned systems also mainly use authorization for write/append access. 
We would need to enforce read access permissions (and see differential privacy and the argument below for the so-called trawl problem).

Self sovereign systems completely decentralize the Id service. 
Fully decentralized systems have a problem with trust and require another component/service  to provide that  e.g. proof of work, stake, community etc.  
These are all known to have scaling or stability and risk challenges and no  convincing solution is as of yet known. 
It is worth remarking that If they are not good enough for currency, why should we trust them for Id systems?

\subsection{Analytics Services}

Systems operators or customers may wish to carry out statistical analyses to audit the proper  operations of identity systems. 
These operations need not be privileged. Differential Privacy is one mechanism to provide privacy with respect to an individual's data in a set. i.e. if someone is doing queries that return aggregates, this only returns results as if the  individual record was not there. 

Applications that use id might also need analytics, for example public health researchers want to look at contact tracing statistics to determine infection rates between users and different classes of user (age, gender etc).

So this might be useful here, but note this has to be for authorized users only (role based access control may be needed), and
there is a limit/quota on number of queries, that is  ``budget'' must be traded off against precision.

Decentralized systems can be coupled with randomness to provide prevention of trawling,  with accurate lookups provided by the model run on users own device and data. But see above for risk problem with decentralized.

For different organizations and  nations, trust assumptions may vary - eg. do you trust a health service provider, government, a bank, a set of individuals, an infrastructure, hardware, operating systems etc Depending on the answer, a different mix of choices from the menu above may be appropriate.

\subsection{Immunity passports}

Here, we provide a brief summary of motivations, concerns and proposals relating to immunity passports. We do not intend this to provide the reader with a complete overview, only to inform them of some of the basics.

The concept of ``immunity passports'' has been considered, and even implemented, by many stakeholders from governments to private companies~\cite{chile_passport}\cite{uk_germany_passport}, as a method to assist in the return to some sort of societal normality. Whether this is to revive an economy, allow workforces to return, or allow the service industry to function at increased capacity, the goal of immunity passports is, at a high level, the same: to reduce the need for social distancing by reducing the risk of transmission of the virus.

It has been suggested the name ``immunity passport'' is misleading \cite{DBLP:journals/corr/abs-2005-11833,10.1001/jama.2020.8102}, instead ``antibody certificates'', ``immunity licenses'' and ``health passports'' have been highlighted as more appropriate names. Moreover, as the scientific evidence on the level and longevity of immunity is fast changing so is the scope of immunity passports. At present however the process is broadly as follows: a test for antibodies is administered and a person is declared immune or not based on the results, the ramifications of having, or not having, an immunity passport is then determined by a government, local authority or employer.

In a similar manner to contact tracing, it is interesting to look at immunity passports from the perspective of identity. An immunity passport can be considered to be a functional identity with at least one mandatory attribute: immunity status, note this is often combined with a photo or other piece of binding information to allow the holder of the passport to “prove” the claim they are making about their immunity status. Herein lies the major concern of such systems however: that they will be inherently discriminatory. Those with immunity will be granted access to post lockdown life, while those without will continue to be restricted. Moreover, the binary nature of the immunity status attribute could highly incentivize a user of the system to act maliciously in order to obtain a passport with ``immune'' status. Consequently, any immunity passporting system must be robust to such behavior. Before we consider some of the proposed systems, we note it has been suggested  [5] that it is not necessary to bind immunity passports to people and thus reduce the level of discrimination. Such a system, however, while decreasing the level of discrimination would likely have far reduced effectiveness. It does highlight an interesting debate in the context of identity however, namely where does the trade-off lie between discrimination and net positive benefit to society?

To the best of our knowledge there have been two academic proposals for immunity passports. First, Eisenstadt et al.~\cite{DBLP:journals/corr/abs-2004-07376}  propose a privacy-preserving scheme that combines W3C-standard ``verifiable credentials''`\cite{sporny2019verifiable}, the ``Solid'' platform \cite{DBLP:conf/www/MansourSHZCGAB16} and a federated consortium blockchain. The authors propose that the hash of each user’s certificate (passport) is stored in a consortium blockchain which is checked each time that an authentication between a user and verifier takes place. Second, Hicks et al.`\cite{DBLP:journals/corr/abs-2005-11833}  present SecureABC: a decentralized, privacy-preserving system, as a solution to the problem. Here a cryptographically signed credential is issued to a user by the healthcare provider and can then be verified by a service provider at any time without the issuer’s (or healthcare provider’s) knowledge.

We also highlight two industry based centralized systems. At a national level, Estonia's system~\cite{estonia_passport}  enables people to share their so-called immunity status with a third-party using a temporary QR-code that is generated after authentication. Commercially, CoronaPass~\cite{coronapass} also propose a centralized immunity passport solution where service providers verify each user passport against a central database. Whilst security and legal measures can be put in place in both these solutions, to deter the central authority from misusing the data they hold, it nonetheless represents an avoidable risk and a central point of failure. Moreover, employing a central party to participate in each authentication risks large-scale user tracking and the possibility of feature creep. The (de)centralized debate in this context is not as nuanced as in the context of contact tracing and therefore we conclude that a decentralized (or blockchain) based approach is likely preferable.

In conclusion, immunity passports offer a method for society to begin to return to normality. Systems have been proposed that provide an effective solution to the passporting problem, and in the case of the academic solutions we have a good understanding of their guarantees and limitations regarding a user’s security and privacy. The implementation of such a system however comes at a cost not only to a person’s privacy but to society; for any such system to have a marked effect in benefiting society it needs to be somewhat discriminatory. As a result we feel a public debate around the level of discrimination we are prepared to tolerate and the associated trade-offs is paramount if such a technology is to be widely implemented.

\section{Measuring efficacy and impact of COVID-19 mitigation methods}
\label{sec:kimpaul}

\subsection{Efficacy of Automated Contact Tracing}
Given the spread of the ongoing SARS-CoV-2 pandemic, automated contact tracing has been suggested as an effective means of containing the spread of the virus while enabling a society to reopen its economy safely. Consequently, a more detailed and rigorous examination of the efficacy of automated contact tracing is required given the distinct difference in the prevalence of this pandemic from the ones in the recent past and the different modes of transmission of the pathogen. 

Manual contact tracing, used more traditionally, has been observed to be effective in previous epidemics caused by the Ebola virus, SARS-CoV and MERS-CoV~\cite{10.1371/journal.pntd.0006762, doi:10.1142/S1793524518500936,BROWNE201533,1551-0018_2018_5_1165,KWOK2019186}. Manual contact tracing is not very effective against pathogens that spread like the influenza virus but is more effective for containing smallpox and SARS-CoV~\cite{10.1371/journal.pone.0000012}. The viral shedding patterns of SARS-CoV and MERS-CoV are similar~\cite{10.1093/cid/civ951,Chowell:2015tn} and show almost no pre-symptomatic transmission~\cite{Fraser6146}, while Ebola is known to be transmitted through the bodily fluids of infected individuals after the onset of symptoms~\cite{REWAR2014444}. On the other hand, influenza shows a significant rate of viral shedding in the pre-symptomatic stage~\cite{Lau:2010tv}. The spreading pattern of SARS-CoV-2 is similar to influenza and quite different from Ebola or SARS-CoV.

To assess the real-world applicability of automated contact tracing, we suggest a model that includes the effects of finite sampling of the population under the assumptions that enrollment in automated contact tracing and reporting on their health condition via the contact tracing service will be voluntary. Not subscribing to the service will not only remove an individual from the pool that is being notified, but it also removes them from the pool of individuals that are reporting while not reporting one's health condition will cause only the latter. In our model, the variabilities in the efficacy of automated contact tracing can be quantified as follows:
\begin{itemize}
    \item Let $N$ be the number of individuals in a population and $f_i$ the fraction of the population that is infected, regardless of whether they know it or not. Therefore, the true number of infected individuals is $f_i N$.
    \item If testing is conducted only when mild or severe symptoms are seen (i.e. excluding testing of asymptomatic cases), the number of confirmed cases is $r_cf_i N$ with $r_c$ being the fraction of the infected that will be confirmed as infected by testing.
    \item We define $f_e$ as the fraction of the population that is enrolled for automated contact tracing and $f_c$ as the fraction of the users that will confirm that they have been diagnosed positive. Hence, the number of individuals that have tested positive, are using automated contact tracing and will confirm that they are sick is $f_cf_er_cf_iN$.
    \item We define $a_c$ as the average number of contacts per person in the period of time $t_0$ who are at risk of being infected due to proximity with a sick individual and is assumed to be greater than 0.
 \end{itemize}
 Using these quantities, we can estimate the number of individuals that can be traced as $f_cf_er_cf_iN \times a_c\times f_e$, i.e. (the number of reported positive tests) $\times$ (the fraction of contacts that will be notified per the report). For automated contact tracing to work effectively, this number should be greater than or equal to the number of individuals that need to be quarantined or isolated since they are now at risk of being infected from coming in contact with a sick person. For the evaluation, we define the following.
 \begin{itemize}
 \setlength\itemsep{0em}
    \item Since $p_t$ is defined as the probability of transmission of infection within the proximity radius $r_0$ being exposed for a time greater than $t_0$, the number of individuals at risk is, at most $p_t f_i N a_c $, i.e., $p_t \times$ (number of contacts of the group of infected individuals).
    \item Finally, we define $f_T$ as the fraction of the individuals at risk of being infected that needs to be successfully quarantined to quell the spread of the pathogen.
 \end{itemize} 
Therefore, the number of individuals that should be quarantined is $f_T p_t f_iNa_c$. For automated contact tracing to slow down the spread of the virus effectively, we have,
\begin{equation}
f_e^2f_cr_cf_iNa_c \geq f_T p_t f_iNa_c.  
\label{eq:need_can}
\end{equation}

Note that $a_c$, the average number of contacts, drops out of the inequality. Hence, the inequality is independent of the population density of the region. This is because eq.~(\ref{eq:need_can}) is in terms of fraction of the population and not the absolute number of individuals. This simply implies that in a region of denser population a larger number of people need to be contacted and quarantined but leaves $f_e$ independent of the population density. Since the right-hand side is the minimum fraction of the population that needs to be traced we arrive at:
\begin{equation}
f^{min}_e = \sqrt{\frac{f_Tp_t}{f_cr_c}}.
\label{eq:min_enroll}
\end{equation}
The fraction $f^{min}_e$ is the minimum fraction of the population that needs to be enrolled in automated contact tracing for it to be effective as a means of slowing down the spread of the pandemic.

Let us examine the limit $p_t=f_c=r_c=1$. This is the limit where every significant contact is assumed to be at risk, everyone who is enrolled in the automated contact tracing program reports sick when tested positive and every sick individual can be successfully identified by testing. Then we arrive at the relation $f^{min}_e = \sqrt{f_T}$. Since $f_T$ is the fraction of contacts that need to be successfully isolated, it can be extracted from the abscissa of Fig. 3 of ref.~\cite{Ferrettieabb6936}. For example, if 100\% of the confirmed infected cases can be isolated, then for a change in the epidemic growth rate by $-0.1$, one needs $f_T\sim 60\%$ and hence $f^{min}_e \sim 77\%$. It is intuitive that $f^{min}_e$ scales as the square root of $f_T$ since both the infected and the contact at risk need to be enrolled and the probability that each are enrolled is $f_e$ leading to $f_T\propto f_e^2$. It gives the threshold which $f^{min}_e$ cannot exceed for any given $f_T$. Several scenarios of the parameter sets are studied in ref.~\cite{2020arXiv200410762K}.

\subsubsection{Assisted contact tracing}

The necessary scale of implementation of automated contact tracing is too large for it to be considered by itself an effective measure to slow down the ongoing pandemic. 
For automated contact tracing to be a viable option, $f^{min}_e$ has to be as low as possible. A closer look at the parameters reveals the following:

\begin{itemize}
\setlength\itemsep{0em}
    \item Both $f_T$ and $p_t$ depend on the dynamics of the disease spread. The fraction of traced cases that need to be quarantined to stop the spread of the disease, $f_T$, can be reduced by extensive monitoring of the disease to make sure sick cases are isolated as soon as possible and their contacts are traced. Even a day or two of delays can increase $f_T$ making automated contact tracing ineffective~\cite{Ferrettieabb6936}.
    \item Variations in $p_t$ can be caused by several factors some of which are controllable. Since $p_t$ depends on the contagiousness of the disease and any protective measures taken against the spread of the infection, $p_t$ can be reduced by measures of limited social distancing, the use of PPE and raising public awareness about the contagiousness of COVID-19. This can pose a significant challenge in densely populated regions and regions with poor living conditions and might lead to the breakdown of the applicability of automated contact tracing.
    \item $f_c$ is somewhat more difficult to control assuming the reporting of those who are confirmed sick is voluntary. This can only be increased by increasing the population's willingness to contribute to automated contact tracing.
    \item $r_c$ is the parameter that is least under control since without very large-scale testing, asymptomatic and mildly symptomatic cases will be difficult to find. This is especially true if the infection can spread by means other than proximity alone as might be the case for SARS-CoV-2~\cite{Santarpia2020.03.23.20039446,doi:10.1056/NEJMc2004973,Guo:2020ww}.
\end{itemize}

Thus we see that a combination of several measures along with a large participation of the population in contact tracing would be the optimal solution for avoiding extensive population-wide social distancing measures and reducing the cost to the economy and well-being of a nation and also allow for greater freedom of movement during a pandemic. In the following section, we discuss future studies on other possible measures that can help in the mitigation of the spread of COVID-19 with particular focus on computational and algorithmic approach leveraging data science and network theory.

\subsection{Intelligent Algorithms, Data-driven Methods and COVID-19}

A pandemic is a population-wide crisis, yet it affects individuals at varying degrees of criticality depending not only on their personal lifestyles but also on the local demographics. It has a unique way of amplifying what history has created as a lingering effect on the current socio-economic trajectories of the immediate neighborhood that any individual lives in. In essence, the effects of and strategies during a pandemic cannot be disentangled from the local socio-economic conditions and historical fluctuations. This renders any large-scale policy implementation without considering the local conditions at the county, state or national level completely ineffective during a pandemic even though the response to it must be population-wide to sufficiently mitigate it. What is imperative looking forward is a means of analyzing localized datasets to understand the curation of mitigation procedures during the onset of a wave of disease spreading. Firstly, data needs to be collected with sufficient granularity to allow for inferences on local prevalence. Secondly, the spread of the disease has to be studied vis-à-vis the local demographics of any neighborhood. Considering the fact that these datasets will be highly multivariate and have very complex correlation pattern obfuscating the causal connections between the driver and the driven, advanced modeling methods and statistical tools are imperative for the understanding of disease spread at a microscopic level of the socio-economic structure of a nation. We have undertaken a three-fold study of various aspects of disease spread and containment that will link them to exit strategies through the mathematical models of curated socio-economic responses and the study of immunity development in a community. These three parts of the study have a deep underlying link which adds to the strength of the analysis that we propose. The three parts can be described as:

\begin{figure}
    \centering
    \includegraphics[trim=10 0 10 20,clip,width=0.49\textwidth]{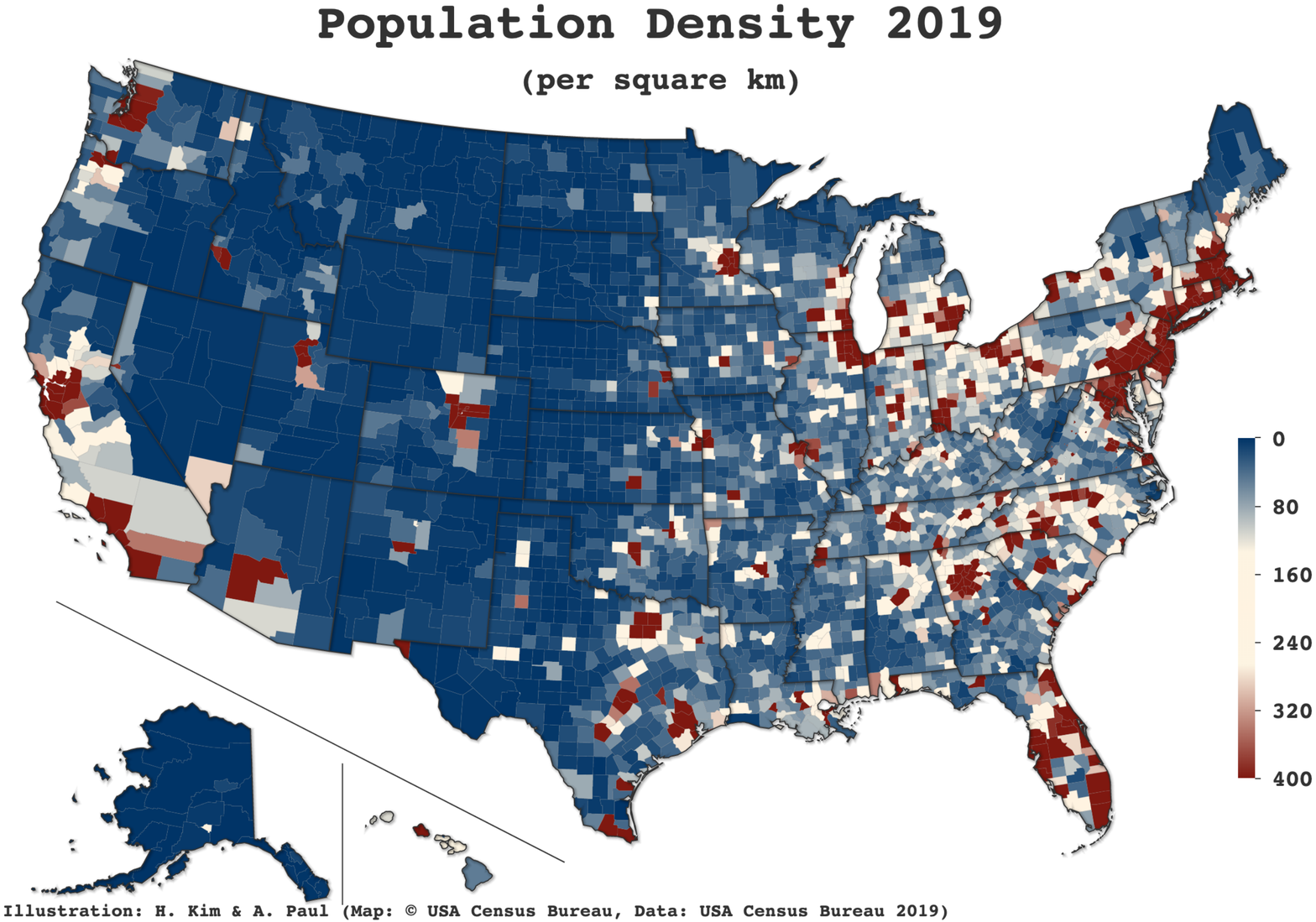}
    \includegraphics[trim=10 0 10 20,clip,width=0.49\textwidth]{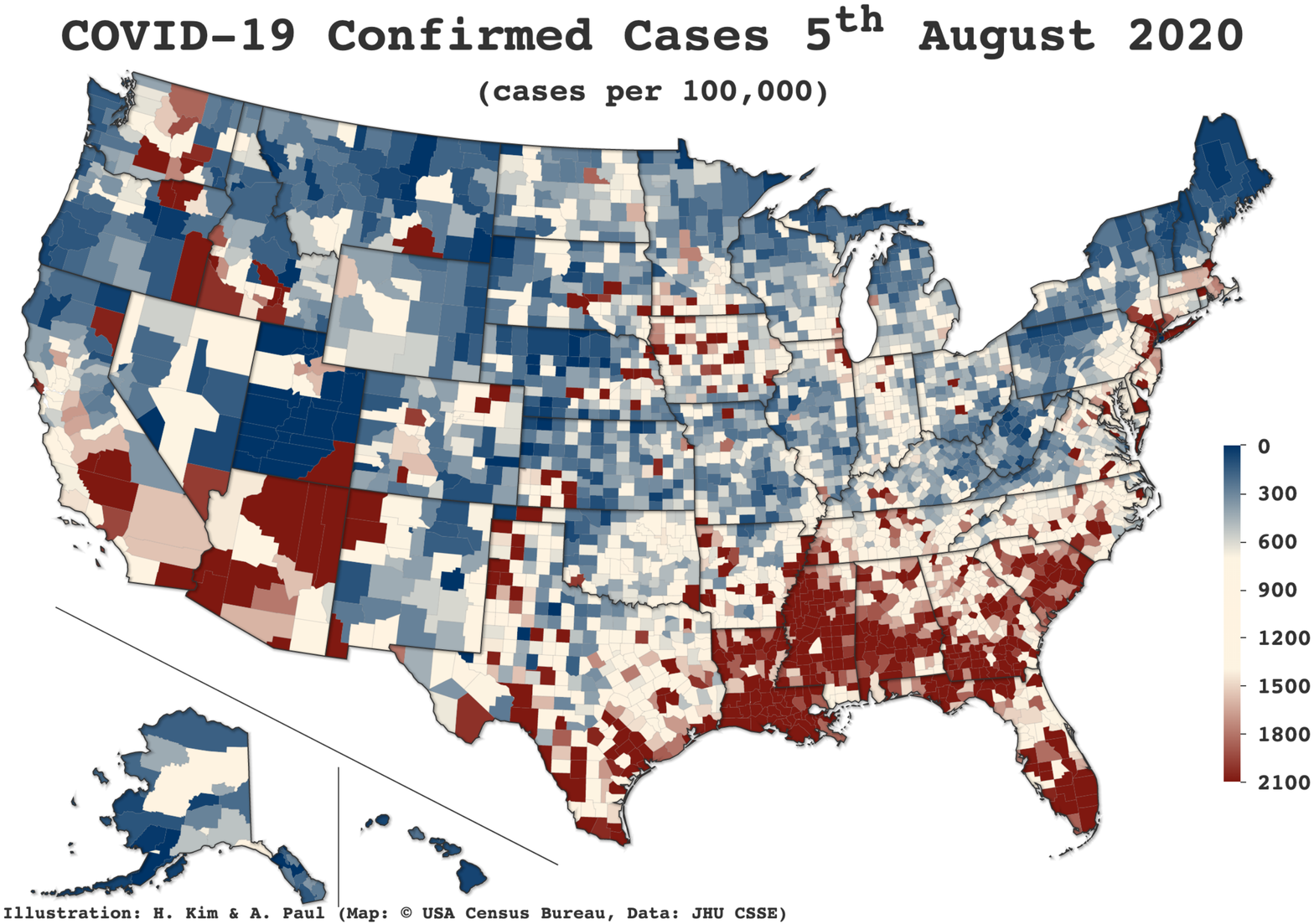}
    \caption{\it Left: distribution of population density in the USA. Right: COVID-19 disease prevalence per 100,000 individuals in each county. The COVID-19 prevalence distribution is quite different from the population density distribution. Plots made with the Highcharts Maps JavaScript library from \href{https://www.highcharts.com/}{Highcharts.com} with a CC BY-NC 3.0 license.}
    \label{fig:maps_US}
\end{figure}

\begin{enumerate}
    \item A study of the correlations between several socio-economic metrics and geospatial demographics at the county level and their correlation with COVID-19 prevalence.
    \item A study of immunity development in a community based on a detailed agent-based model and the training of a machine learning algorithm to probabilistically assess and categorize the immunity development in individuals.
    \item A network-based model of urban areas to understand curated closures of the commercial and industrial sectors to find an optimal level between uncontrolled disease spread and damages to the economy.
\end{enumerate}

All three topics tie into a common goal: understanding the socio-economic conditions that affect the spreading of the pandemic and devising effective exit strategies and mitigating policies that can allow for the sustenance of economy while allowing for lower footprint of a pandemic in terms of human lives lost and perturbations to the social norm. These strategies will be augmented by an understanding of immunity development which is crucial for increasing mobility within a population and will build tools that will allow for algorithmic assessments to aid in prioritizing clinical testing for immunity. In what follows, we give some details of the ideas that we are exploring in these studies.

\subsubsection{Socio-Economic conditions and COVID-19}\label{subsec:socio-metric}

As COVID-19 spreads by contact and proximity, it is natural to assume that the pandemic will have its worst effects in regions that have the highest density of population and the largest mobility within the population. 
This effect can even be seen in several nations where, on the surface,  the worst-hit regions were the most densely populated ones. 
However, Fig.~\ref{fig:maps_US} tells a very different story. 
In the USA, there seems to be not a very tight correlation nationally between the population density of a region and the prevalence of COVID-19. 
Unexpected correlations between the prevalence of COVID-19 and socio-economic metrics can appear especially in rural areas. 
Correlation of COVID-19 prevalence with socio-economic metrics like mobility, unemployment, poverty, the fraction of population that are migrants, types of professions undertaken, etc. that can be found in the census data at the county level can be studied. To understand the correlations better, we are using machine learning tools like boosted decision trees along with feature importance measures like the Shapley score to understand how the socio-economic metrics affect the disease spread.

\subsubsection{Algorithmic assessment of immunity to COVID-19}\label{subsec:immunity}

In another work we study the disease spread using an agent-based model~\cite{Cliff2018,Zachreson2018} and develop a machine learning algorithm that will be able to identify immune individuals in a data-driven manner. The study requires a simulation of an agent-based model to create a simulated data-set of immune agents starting with a small number of sick agents. To set up the rules for the agent-based model there has to be an understanding of how the disease propagates in real communities affected by the disease. 
This requires the gathering knowledge from emerging clinical studies of COVID-19 affected communities. 
The immunity detection algorithm is trained using the simulated data-set generated by the agent-based model. 
This algorithm will evolve into a semi-supervised machine learning algorithm that will learn the optimal values of the parameters necessary for inferring on immunity development in individuals. 
Data obtained in \ref{subsec:socio-metric}, related to local infection prevalence, duration of pre-symptomatic and symptomatic stages of the infection, demographic data on infected individuals, assessment of the prevalence of asymptomatic cases helps in the design of the simulation which will be used to train the immunity detection algorithm. 

From this we will gain a data-driven understanding of immunity development in a population. This will act as a starting point when real data is made available with the deployment of anti-body tests. An assessment of individual immunity based on prior exposure, underlying health conditions, local disease spread data and mobility history will allow individuals to be aware of possible immunity development and allow any institution (within the healthcare system or otherwise) to prioritize immunity tests based on algorithmic assessment. This will allow for better assessment of the necessity for quarantining individuals and will ease the requirement for population-wide stay-at-home orders, and can ultimately be a reference to making policy decision for reopening for COVID-19 as well as future epidemics.

\subsubsection{A network-based analysis for optimized commerce management}

Exit strategies during an ebbing pandemic requires special caution so as to not trigger its resurrection as is being seen in several parts of the world now. A key component is understanding the fraction of the various commercial sectors that can be opened to sustain the flow of the economy while optimizing the social contact within the population. There are certain businesses that have a higher footfall and hence act as transmission hubs like restaurants and supermarkets. 
If a fraction of these hubs is closed intermittently, it can reduce the effective pathway for disease propagation, hence, slowing down the disease spread while it allows keeping the businesses at a sustainable level. However, this requires accurate large-scale prediction about what fraction of different business sectors need to be closed. 
To this end, we are developing multi-layered network models~\cite{Eubank:2004tj, Kim_2012, PhysRevLett.117.208301} to represent businesses as nodes with their interdependence as links in one layer and the pathways of disease spread in another layer. 
Utilizing the business interaction network on the first layer, we investigate how the partial closure of certain businesses affect other businesses that depend on them. Then, we analyze mobility patterns on the business network by using mobility data collected from Google and identify hubs associated with high mobility in an urban area. 
We adopt the agent-based model discussed in \ref{subsec:immunity} to simulate the mobility data and tune it with the fraction of open business on the first layer to reduce the mobility at the hubs. This enables the analysis of how the intervention on the business network propagates and influence the spread of disease across the social interaction network on the second layer~\cite{NUSS201614,PhysRevLett.88.228102}.
%
%

\section{Concluding Remarks}
\label{sec:conclusion}

Since the workshop, the pandemic has significantly worsened, and the policy responses to upsurges in cases has varied widely from nation to nation, and even state to state. As of at September $27^{th}$, the global cases are approaching $33$ million, and sadly deaths now stand at $995,583$ with a difficult Northern Hemisphere winter to come. The impact of restrictions on personal liberty and economic activity continue to be manifested in deep recessions, significant changes in patterns of travel, and increasingly in civil unrest. The importance of seeking ‘smart solutions’ to manage the pandemic with minimal disruption is even more pressing than when the workshop was convened. We hope that the results presented here will inform both researchers and policy makers.

It is clear that there is much left to be understood regarding the interplay between the network of contacts and the progression of the epidemic. As was explained in Sections 2-5 the precise structure of the contact network. However, detailed information about the network structure is scarce and most of the modeling approaches have been based on coarse-grained data about the network structure. An exception to this rule is the case in which data from track and tracing apps is taken into account, but even here, the data includes only partial information and needs to satisfy the privacy agreement set up with its customers. Due the adaptive nature of the human response to COVID-19, making predictions regarding the unfolding of the epidemics is hazardous. In particular, in the countries where a partial containment of the spread of the epidemic has been achieved, there remain uncertainties over the evolution of the disease and policy makers are faced with a delicate situation  where the pandemic can easily spiral out of control.  

The computational challenges that underlie these models represent an opportunity for a novel approach to studying the effects of the network structure. Most of the analytical approaches require simplification of the network structure, and it is entirely possible that the glossed over fine details of this structure could have a dramatically alter the conclusions obtained. This situation could be viewed as similar to that presented by the atomic hypothesis  nearly a century and a half ago. That problem was brilliantly solved by the body of work now known as statistical physics. Analogous approaches have been adopted in network science, and it is possible that equally consequential advances are just around the corner, with obvious implications for more effective containment measures and immunization policies.

A much vaunted approach to managing the disease is automated contact tracing. It is also clear that in this endeavor there is much to be gained by further study and innovation, for example, a more complete understanding of the contact network though which COVID-19 spreads. Reliable data will be essential for the application of advanced network science techniques to predict the spread to the disease.

At the heart of such approaches lies the balance between personal privacy and the health of the population at large. Advances in the trustworthiness of such systems described in Sections 6 and 7 could be pivotal in driving the adoption and efficacy of technological approaches to controlling the pandemic. 

Finally it is worth noting that the socio-economic profile of the impact of COVID has laid bare structural challenges in developed, and developing societies. Hitherto the world has largely been urbanized or urbanizing, and anecdotally at least this has paused. If, as some believe, this pandemic heralds a period of recurring pandemics, it seems likely that a new vision for social and economic progress will be needed. It is, of course, not the place of scientists to dictate what that vision might be: it is a question for all of society. But we hope the insights and developments presented in this article will contribute positively to whatever vision emerges. 

\ack
I.Z. Kiss and F. Di Lauro acknowledge support from the Leverhulme Trust for the Research Project Grant RPG2017-370.

\section*{References}
\bibliographystyle{ieeetr}
\bibliography{AsuCv19Perspectives}
\end{document}